\documentclass{jfm}
\usepackage{graphicx}
\usepackage{lineno}
\usepackage{color}
\usepackage{amsmath}
\usepackage{amssymb}
\usepackage{mathrsfs}
\usepackage{gensymb}
\usepackage{verbatim}
\usepackage{cleveref}
\usepackage{mathtools}
\usepackage[normalem]{ulem}

\shorttitle{Generation of attached Langmuir circulations}
\shortauthor{C. Yan, J. C. McWilliams and M. Chamecki } 

\title{Generation of attached Langmuir circulations by a suspended macroalgal farm}

\author
 {
  Chao Yan,
  James C. McWilliams\aff{}
  \and
  Marcelo Chamecki\aff{}
  \corresp{\email{chamecki@ucla.edu}}
  }

\affiliation
{
\aff{}
Department of Atmospheric and Oceanic Sciences, University of California, Los Angeles, CA 90025, USA
}

\begin{document}

\maketitle

\begin{abstract}
In this study, we focus on Langmuir turbulence in the deep ocean with the presence of a large macroalgal farm using a Large Eddy Simulation method. The wave-current interactions are modelled by solving the wave-averaged equations. The hydrodynamic process over the farm is found to drive a persistent flow pattern similar to Langmuir circulations but is locked in space across the farm. These secondary circulations are generated because the cross-stream shear produced by the rows of canopy elements leads to a steady vertical vorticity field, which is then rotated to the downstream direction under the effect of vortex force. Since the driving mechanism is similar to the Craik-Leibovich type 2 instability theory, these secondary circulations are also termed as attached Langmuir circulations. We then apply a triple decomposition on the flow field to unveil the underlying kinematics and energy transfer between the mean flow, the secondary flow resulting from the farm drag, and the transient eddies. Flow visualizations and statistics suggest that the attached Langmuir circulations result from the adjustment of the upper ocean mixed layer to the macroalgal farm, and they will weaken (if not disappear) when the flow reaches an equilibrium state within the farm. The triple-decomposed energy budgets reveal that the energy of the secondary flow is transferred from the mean flow under the action of canopy drag, while the transient eddies feed on wave energy transferred by the Stokes drift and energy conversion from the secondary flow.
\end{abstract}


\section{Introduction}\label{sec:1}

Macroalgae, also known as seaweeds, are an important component in temperate marine ecosystems \citep{Dayton:1985ARES, Schiel:2015}. Providing shelter, food and protection for many species of marine living creatures, macroalgae play a paramount role in preserving biodiversity and promoting sustainable aquaculture production. Macroalgal forest harvesting also contributes enormously to various applications, such as remediation of eutrophication pollution, biofuel production, food and pharmaceutical processing, etc. The desire to increase the productivity of aquaculture spurs the growing need for aquafarm development in the ocean, where the canopy grows near the surface and is supported by a floating structure \citep{Troell:2009, Stevens:2011aei}. The macroalage canopy alters the surrounding flow conditions by dampening the currents and wave motions \citep{Rosman:2007JGR}. These flow modifications have profound implications for the nutrient uptake and associated processes of sedimentation and recruitment \citep{Duggins:1990jembe, Plew:2011aei}. Therefore, understanding and quantifying the diverse hydrodynamic processes that occur in the presence of macroalgal farms is essential in evaluating and designing optimal farm configurations, as well as assessing their environmental impacts.

From a hydrodynamics perspective, aquatic vegetation can be classified as submerged, emergent, or suspended based on its growth form. Submerged and emergent vegetation are attached to the bottom floor, and occupy a fraction or all of the water depth. The flow structures and mass transport over such canopies have been well documented \citep{Nepf:2012ARFM, Nepf:2012JHR, Yan:2017efm}. Particular attention has been given to the shear layer turbulence at the canopy top (for submerged canopy), which prompts the generation of canopy-scale coherent structures that dominate the momentum and scalar exchanges between the canopy and the free flow above. Suspended canopies, such as the macroalgal farm considered here, extend downward from the surface and occupy the upper part of the water body \citep{Plew:2005joe, Plew:2006efm, Stevens:2011aei}. The flow and canopy interactions for this configuration remain less explored as compared to the bottom-mounted counterpart \citep{Stevens2019fmars}.

For suspended vegetation, the vertical discontinuity in drag beneath the canopy also leads to a shear layer, which penetrates a finite distance into the canopy and mediates the turbulent exchanges between the canopy and the underlying flow \citep{Plew:2011jhe}. Through laboratory experiments of suspended canopies in shallow waters, \citet{Plew:2011jhe} concluded that the additional bottom boundary layer (BBL) associated with the ocean floor affects the penetration of the shear layer into the suspended canopy. Based on the measurements of \citet{Plew:2011jhe}, \citet{Huai2012awr} proposed a simple analytical model for the vertical profile of streamwise velocity. While these studies focus on flow over uniform canopies (i.e. essentially infinite size), where the flow has been fully adjusted to the canopy, common aquaculture structures are of finite size and the corresponding canopy flow displays distinct spatial distribution patterns.

The finite dimensions and spatial arrangement of the suspended canopy lead to flow patterns different from the fully developed scenario \citep{Tseung:2016efm}. According to \citet{Tseung:2016efm} and \citet{Zhao2017AWR}, the flow over a suspended canopy of finite size is similar to the terrestrial flow over forest patches \citep{Belcher2003jfm}, and it can be divided into four zones of distinct mean flow behaviour in the downstream direction: (i) the upstream adjustment zone, (ii) the transition zone, (iii) the fully developed zone, and (iv) the wake zone. The distance over which the velocity profile reaches a fully developed state is affected by canopy geometry (e.g. plant density and stem diameter) \citep{Rosman2010lo}. \citet{Zhou:2019jfm} examined the effect of a circular patch of suspended canopy on the mean flow dynamics in deep water, and found out that the patch geometry poses another impact on the adjustment of flow pathways. In the light of these studies, we are motivated by the water flow over an aquaculture farm of finite size in deep ocean, and seek to explore how the ocean mixed layer (OML) evolves as it approaches and flows over the farm under typical ocean conditions.

In the marine environment, ocean waves have a profound influence on the water flow and the exchange of nutrients between kelp forests and ambient water. In many studies, this effect is characterized in terms of the Stokes drift \citep{Gaylord:2007lo, Rosman:2007JGR}, which refers to the net motion of fluid parcels in the direction of wave propagation that arises from the unclosed orbital motions for finite amplitude waves \citep{Monismith:2004lo}. \citet{Rosman:2007JGR} explored the effects of giant kelp forests on ocean flows through a field experiment at the coast of Santa Cruz, California. They highlighted the importance of the Stokes drift in cross-shore transport within the kelp canopy. \citet{Rosman:2013lo} conducted experiments at a scaled laboratory flume to examine the interaction of surface waves and currents with kelp forests, and concluded that these interactions must be taken into account when modeling flow and transport within kelp forests. 

One of the distinct features widely observed in the upper ocean is the presence of Langmuir circulations, which consists of counter-rotating vortices near the ocean surface roughly aligned with the wind direction \citep{Thorpe2004ARFM}. It is well accepted that the Langmuir circulations are generated by the interaction between the wind-driven shear current and the Stokes drift velocity induced by the surface gravity waves through the the Craik-Leibovich (CL) type 2 instability \citep{Craik:1977JFM, Leibovich1983arfm}. The associated ocean flows are referred to as Langmuir turbulence \citep{McWiliams:1997jfm}, which can be numerically modelled by adding a vortex force into the momentum equation without the need to resolve the surface gravity waves \citep{Skyllingstad:1995jgr, McWiliams:1997jfm, Yang:2015jgr, Chamecki2019RoG}. The increased level of turbulence intensity promoted by Langmuir circulations is expected to affect the supply and uptake of nutrients within the marine ecosystem \citep{Barton2014}. 

In this study, we use a fine-scale Large Eddy Simulation (LES) model to explore the development of an OML in the presence of a large macroalgal farm under typical current and wave regimes. The main goal of the present work is to characterize the hydrodynamics around a macroalgal suspended farm and advance our understanding of canopy flows in the ocean. We assume that the ocean is deep enough so that the flow is free from the complexities of BBL. Section \ref{sec:2} describes the numerical approach for modeling oceanic boundary layer flow over a macroalgae canopy. A triple decomposition strategy is used to separate the flow field into the contributions due to mean flow, secondary flow resulting from the farm drag, and transient fluctuations. Section \ref{sec:3} describes the main characteristics of the flow field and the emergence of persistent flow structures termed ``attached Langmuir circulations''. 
Section \ref{sec:mechanism} discusses the underlying mechanism of generation of attached Langmuir circulations, and characterizes their spatial development. Section \ref{sec:energetics} describes the energy conversion among the three components of the flow field. Conclusions are drawn in section \ref{sec:conclusion}.

\section{Methods}\label{sec:2}
\subsection{Mathematical model}

For the past three decades, the LES technique has been widely adopted to study turbulence in the OML. Detailed discussion of the LES framework and assumptions underpinning its applicability can be found in the review paper by \citet{Chamecki2019RoG}. In the present work, the dynamics of Langmuir turbulence in the presence of a macroalgae canopy are captured using the LES method by solving the wave-averaged equations described by \citet{McWiliams:1997jfm}. This mathematical model is built upon the original Craik-Leibovich equations \citep{Craik:1976JFM} with the inclusion of planetary rotation and Stokes drift advection of scalar fields,
\begin{equation} \label{eq:continuity}
    \nabla \cdot {\widetilde{\boldsymbol{u}}}=0, 
\end{equation}
\begin{equation} \label{eq:momentum}
    \frac{\partial{\widetilde{\boldsymbol{u}}}}{\partial{t}}+{\widetilde{\boldsymbol{u}}} \cdot \nabla{\widetilde{\boldsymbol{u}}} =-\nabla{\Pi}-f\boldsymbol{e_z}\times\left(\widetilde{\boldsymbol{u}}+\boldsymbol{u}_s-\boldsymbol{u}_g\right)+\boldsymbol{u}_s \times \widetilde{\boldsymbol{\zeta}}+\left(1-\frac{\widetilde{\rho}}{\rho_0}\right)g\boldsymbol{e_z}+\nabla \cdot {\mathsfbi{\boldsymbol{\tau}}^d}-\boldsymbol{F}_D,
\end{equation}
\begin{equation} \label{eq:rho}
    \frac{\partial{\widetilde{\rho}}}{\partial{t}}+ \left(\widetilde{\boldsymbol{u}}+\boldsymbol{u}_s\right) \cdot \nabla \widetilde{\rho}=\nabla \cdot {\boldsymbol{\tau}_\rho},
\end{equation}
Here, the tilde indicates grid-filtered variables, $\widetilde{\rho}$ is the filtered seawater density, $\rho_0$ is the reference density, $\Pi$ is the generalized pressure, $f$ is the Coriolis frequency, $g=9.81\ \mathrm{m\ s^{-2}}$ is the gravitational acceleration, $\boldsymbol{e_z}$ is the unit vector in the vertical direction, and ${\widetilde{\boldsymbol{u}}}=({\widetilde{u}, \widetilde{v}, \widetilde{w}})$ is the velocity vector  represented in the Cartesian coordinate system $\boldsymbol{x}=({x}, {y}, {z})$, with $x$, $y$, and $z$ being the downstream, cross-stream, and vertical directions, respectively. The vertical coordinate is defined positive upward with $z=0$ at the ocean surface. The geostrophic current $\boldsymbol{u}_{g}=(u_g, 0, 0)$ is driven by an external mean pressure gradient force with magnitude $fu_g$ applied in the $y$-direction. The canopy is treated as a source of flow resistance, and its effect is accounted for by adding a drag force $\boldsymbol{F}_D$ to the momentum equation. 

In \eqref{eq:momentum} and \eqref{eq:rho}, ${\boldsymbol{\tau}^d}$ is the deviatoric part of the subgrid-scale (SGS) stress tensor $\boldsymbol{\tau} =\widetilde{\boldsymbol{u}}\widetilde{\boldsymbol{u}}-\widetilde{\boldsymbol{u}\boldsymbol{u}}$, and ${\boldsymbol{\tau}_\rho} = \widetilde{\boldsymbol{u}}\widetilde{\rho}-\widetilde{\boldsymbol{u}\rho}$ is the SGS buoyancy flux. We assume that the changes in the seawater density $\rho$ are caused by the varying potential temperature $\theta$, and these two variables are linearly related by $\rho=\rho_0[1-\alpha(\theta-\theta_0)]$, where $\alpha=2\times10^{-4}\ \mathrm{K}^{-1}$ is the thermal expansion coefficient, and $\theta_0$ is the reference potential temperature at which $\rho_0$ is measured. The SGS stress tensor is modeled using the Lagrangian scale-dependent dynamic Smagorinsky SGS model \citep{Bou-Zeid:2005pof}. Then, the SGS buoyancy flux is parameterized using an eddy diffusivity closure with a prescribed value of SGS Prandtl number $\Pran_t=0.4$. The viscous force is assumed to be negligible for the high-Reynolds number flows considered in the present study.

The Stokes drift $\boldsymbol{u}_{s}$ induced by surface gravity waves is imposed in the governing equations to reflect the time-averaged effects of the wave field on the oceanic turbulence, since the surface wave motions are not explicitly resolved in our simulations. The third term on the right-hand-side (RHS) of \eqref{eq:momentum} is the CL vortex force $\boldsymbol{u}_s \times \widetilde{\boldsymbol{\zeta}}$ (here $\widetilde{\boldsymbol{\zeta}}=\nabla \times \widetilde{\boldsymbol{u}}$ is the vorticity field), which represents the interaction of wind-driven turbulence and surface gravity waves. For simplicity, we only consider a steady monochromatic wave. Assuming that the surface gravity wave propagates along the mean wind direction (i.e. the $x$-direction), the Stoke drift velocity reduces to $\boldsymbol{u}_{s}=(u_s(z), 0, 0)$, where $u_s$ is given by:
\begin{equation} \label{eq:stokes drift}
    u_s={U_s}e^{2kz}
\end{equation}
in which $k$ is the wavenumber and $U_s$ is the wave-induced Stokes drift at the surface. Then, the vortex force $\boldsymbol{u}_s \times \widetilde{\boldsymbol{\zeta}}$ reduces to $(0, -u_s \widetilde{\zeta}_z, u_s \widetilde{\zeta}_y)$. Note that the presence of the canopy can attenuate the waves and impact the Stokes drift profile \citep{Rosman:2013lo}. Based on the approach developed by \citet{Dalrymple1984}, we have estimated the effects of canopy drag on the surface waves for the specific canopy and wave parameters used in this study and found only a small attenuation of about $3\%$ in wave amplitude and $6\%$ in the magnitude of the Stokes drift (see appendix \ref{appB}). These estimates are consistent with those obtained in flume measurements by \citet{Rosman:2013lo}. For the sake of simplicity, we neglect wave attenuation in this study.

Finally, there is evidence suggesting that surface waves can induce a mean current in the direction of the wave propagation within aquatic canopies \citep{Luhar2010JGR, Luhar2013JGR, Abdolahpour:2017JGR, Chen2019AWR, Rooijen2020JGR}. This wave-induced current is caused mainly by the reduction of the wave orbital velocity within the canopy, and inclusion in our model would require explicitly resolving the surface waves. We used the empirical results in the literature \citep[see][]{Abdolahpour:2017JGR, Chen2019AWR} to estimate the maximum magnitude of this wave-induced current for the suspended farm simulated here, and found out that it is a reasonably small fraction (about 20\%) of the steady geostrophic current $u_g$ imposed in our simulations. Thus, we expect that the overall effects of this wave-induced current to be small, and we neglect them in adopting a wave-averaged approach.


\subsection{Numerical representation of macroalgal farm}
 
\begin{figure}
    \centerline{\includegraphics[width=1.0\textwidth]{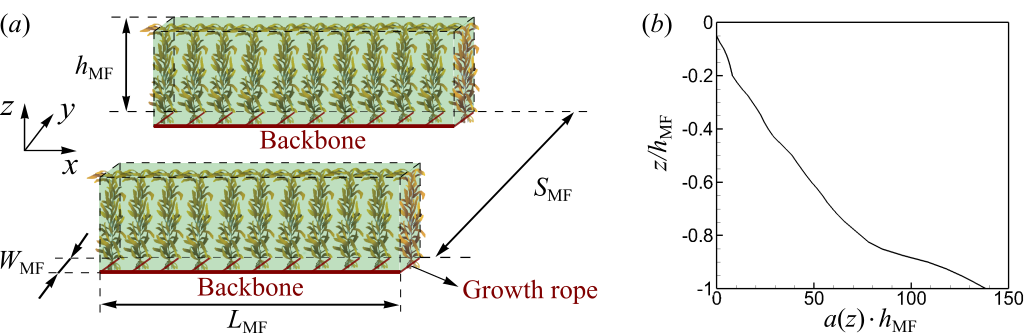}}
    \caption{Schematic of the spatial morphology of the suspended macroalgae farm: (\textit{a}) spatial arrangement of the macroalgal farm; (\textit{b}) frond area density profile for each macroalgae row, $a(z)$, normalized by the canopy height $h_\mathrm{MF}$.}
\label{fig:morphology}
\end{figure}

For the cultivation of macroalgae, the aquaculture structures being deployed in the open ocean are varied, but common practice is to suspend seeded materials from surface buoys and mooring structures \citep{Charrier:2018}. One possible configuration for the cultivation strategy for the macroalgal of interest (giant kelp) is shown in figure \ref{fig:morphology}\textit{a}. The macroalgal farm comprises parallel lines of seeded growing ropes with a length of $W_\mathrm{MF}=8$ m coiled around a backbone (or longline). Each backbone line, with a length $L_\mathrm{MF}$, is anchored at each end and connected to surface buoys (not shown). Each macroalgae consists of 8 fronds with an average length $h_\mathrm{MF}=19$m, which are assumed to be in an upright posture by virtue of the buoyancy provided by the gas-filled floats (called pneumatocysts). The lateral spacing between two adjacent rows of canopy elements is $S_\mathrm{MF}=26$m. 

The frond surface area of the cultivated macroalgae species is obtained by conversion of vertically-resolved algal biomass generated from a macroalgal growth model (Christina Frieder, personal communication) using allometric relationships \citep{Fram2008LO}. To simplify the numerical modeling, the frond surface area is redistributed uniformly within each canopy row in the horizontal directions, while the spatial arrangement of the row structure is resolved in the simulation. The fraction occupied by the macroalgae elements has a total foliage area density (FAD) profile denoted as $a(z)$, which is shown in figure \ref{fig:morphology}\textit{b}. FAD is the total (one-sided) frond surface area per unit volume of space ($\mathrm{m^{-1}}$), without explicit differentiation among blades, fronds, and stipes, etc. Since our main focus here is to examine the adjustment of OML as it flows over the farm, canopy parameters such as $a(z)$, $h_\mathrm{MF}$, $S_\mathrm{MF}$ are kept constant (the only exception being the length $L_\mathrm{MF}$) and a sensitivity study to farm design is beyond the scope of this study. 

The drag per unit mass $\boldsymbol{F}_D$ in \eqref{eq:momentum} represents the effect of the canopy as a momentum sink for the flow field, and it is parameterized as \citep{Shaw:1992BLM, Pan2014jfm},
\begin{equation} \label{eq:drag}
    \boldsymbol{F}_D=\frac{1}{2}C_D a(z)\mathsfbi{P} \boldsymbol{\cdot}|\widetilde{\boldsymbol{u}}|\widetilde{\boldsymbol{u}}
\end{equation}
in which $C_D$ is the drag coefficient and $|\widetilde{\boldsymbol{u}}|$ is the magnitude of the resolved velocity vector. For the sake of simplicity, the tilde symbols used to denote resolved variables are omitted hereafter. The coefficient tensor $\mathsfbi{P} = P_x\boldsymbol{e_x}\boldsymbol{e_x} + P_x\boldsymbol{e_y}\boldsymbol{e_y} + P_z\boldsymbol{e_z}\boldsymbol{e_z}$ is employed here to account for the projection of total foliage area onto the orthogonal planes with normal in each one of the Cartesian directions. Note that the expression for $\mathsfbi{P}$ involves the dyadic products of the standard basis vectors $\boldsymbol{e_x}$, $\boldsymbol{e_y}$, and $\boldsymbol{e_z}$, so that $\mathsfbi{P}$ is also a second-order tensor. This projection operation is commonly used for terrestrial canopies \citep{LEGG1979, Aylor:2001JAM, Pan2014jfm}, and the coefficients $P_x$, $P_y$, and $P_z$ depend on the geometry of the canopy and thus on the specific details of each plant species \citep{Aylor:2001JAM}. In the absence of observational data to specify these coefficients, we make the assumption of isotropic distribution of FAD (e.g. the fraction of FAD projected towards each direction is always the same), which corresponds to $P_x = P_y = P_z = 1/2$. 

The drag coefficient $C_D$ is a key input parameter in the drag model \eqref{eq:drag} that can affect the accuracy for the prediction of turbulence statistics \citep{Pinard2001JAM}. Generally, $C_D$ is estimated from the reduced momentum balance based on experimental measurements, where large uncertainty exists depending on the formulations of the momentum equation being used \citep{Cescatti2004AFM, Pan2016BLM} and quality of measurements \citep{Pinard2001JAM, Marcolla2003BLM}. Many numerical studies of atmospheric boundary layer flows used a height-averaged $C_D$ of constant value for terretrial canopies \citep{Shaw:1992BLM, Dupont:2008AFM, Finnigan2009JFM}. For flexible canopy like the macroalgae, the canopy elements can bend back and forth with the moving water, leading to reduced fluid drag relative to the rigid and upright vegetation \citep{Boller2006jeb, Luhar2011lo}. \citet{Pan2014jfm} introduced a velocity-dependent $C_D$ in their LES study to account for the reconfiguration of the flexible cornfield in response to the surrounding flow \citep{Vogel:1989}. However, giant kelp elements do not bend with the flowing water in the same way as many terrestrial plants or seagrasses do, because they possess many gas-filled floats that can keep the fronds upward to the surface via buoyancy forces \citep{Koehl:1977lo, Henderson2019ce}. In our LES cases, we use the value of $C_D = 0.0148$ reported in the experimental study of \citet{Utter:1996}, which measured the drag coefficient on Macrocystis pyrifera fronds by towing a single plant from a boat in a field experiment. It should be noted that \citet{Utter:1996} modeled the canopy drag by a power law of the local velocity with an exponent of 1.6  to account for the drag reduction resulting from plant reconfiguration, while we assume the relationship between these two variables to be quadratic (equation \eqref{eq:drag}). 

Apart from the fluid drag force, the macroalgae plants are subjected to elastic and buoyant forces, both of which act to resist bending. The subtle balance among these forces determines the posture of macroalgae elements \citep{Luhar2011lo, Henderson2019ce}. Estimates given in appendix \ref{appA} show that  kelp stipes remain approximately upright in the flow, except for an oscillatory motion with amplitude comparable to the wave orbital displacement. In fact, this assumption is implicit in the parametric model for the drag force \eqref{eq:drag}: the wave orbital velocity is not included in the drag calculation, implying that the macroalgae plants oscillate with the wave orbital velocity (note that this assumption is consistent with the idea that the macroalgae canopy does not impact the waves).

\subsection{Numerical scheme}
The present LES framework employs a Cartesian grid using a vertically staggered arrangement, with the horizontal velocity components, pressure and potential temperature $(u, v, p, \theta)$ defined at the cell center, while the vertical velocity component $(w)$ is stored at the cell face. Spatial derivatives in the horizontal directions are treated with pseudo-spectral differentiation, while the derivatives in the vertical direction are discretized using a second-order centered-difference scheme. Aliasing errors associated with the non-linear terms are removed via padding based on the 3/2 rule.  Time advancement is performed using the fully explicit second-order accurate Adams-Bashforth scheme. The numerical code has been validated against the LES study of \citet{McWiliams:1997jfm} for Langmuir turbulence in deep ocean by \citet{Yang:2015jgr}.

\begin{figure}
    \centerline{\includegraphics[width=0.8\textwidth]{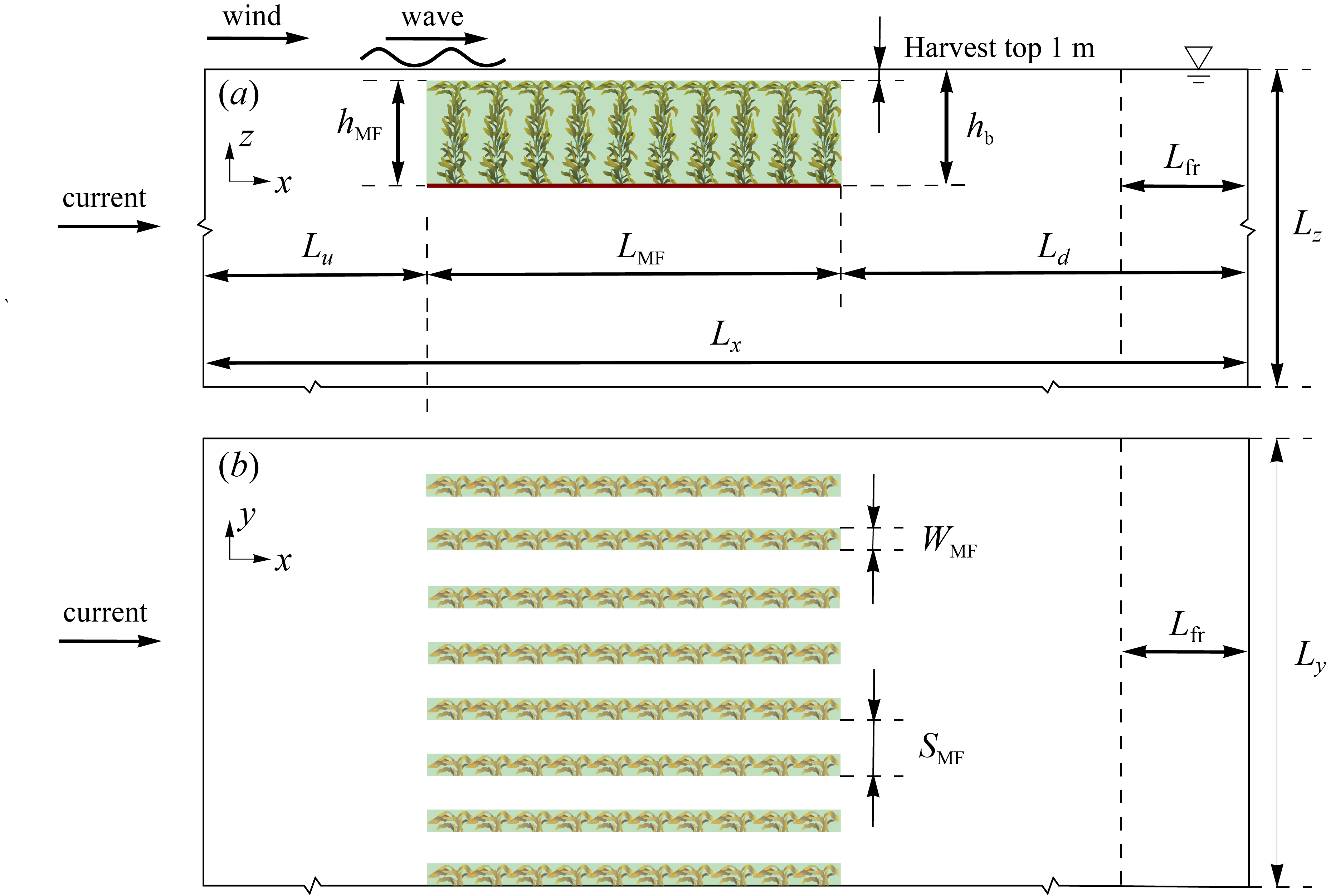}}
    \caption{Sketch of the LES computational model for Langmuir turbulence with the presence of suspended macroalgae farm in deep ocean: (\textit{a}) side view and (\textit{b}) plan view. A fringe region of length $L_\mathrm{fr}$ towards the end of the domain is used to force the velocity and potential temperature back to the inflow, so periodic conditions are satisfied in the horizontal plane.}
\label{fig:domain}
\end{figure} 

The LES domain with dimensions of $L_x \times L_y \times L_z$ is shown in figure \ref{fig:domain}. For clarity, the origin of the coordinate system is defined at the leading edge of the farm in the central longitudinal plane, and the $z$-axis is pointing upward. The top boundary is specified as a non-deforming surface exposed to wind shear stress. A sponge layer is imposed within the bottom 20\% of the domain to damp out fluctuations of velocity and temperature, thus avoiding the reflection of the internal gravity waves.

The backbone line is at a depth $h_\mathrm{b}=20$m below the surface while the canopy height is $h_\mathrm{MF}=19$m, leaving a canopy-free layer at the top 1m near the ocean surface to represent typical harvest practices. A domain depth of $L_z=6h_\mathrm{b}$ is chosen to avoid the interference with the bottom boundary condition as the flow is deflected below the canopy. The cross-stream domain size $L_y=8S_\mathrm{MF}$ is tailored to encompass $N=8$ parallel rows of macroalgae elements, the longitudinal axes of which are aligned in the downstream direction. Periodic boundary conditions are imposed in the horizontal directions, which will enable us to exclude the complexities brought by the limited width of the farm. The inlet is positioned $L_u=7.5h_\mathrm{b}$ upwind from the farm leading edge, and the outlet is at a distance $L_d=12.5h_\mathrm{b}$ downstream of the farm trailing edge. Thus, the domain size in the downstream direction is $L_x = L_\mathrm{MF}+20h_\mathrm{b}$.

A fringe region of length $L_\mathrm{fr}=5h_\mathrm{b}$ is used at the end of the domain (see figure \ref{fig:domain}) to enable simulations of spatially evolving boundary layer flows in a periodic domain using pseudo-spectral numerics \citep{Stevens2014RE}. Specifically, the inflow turbulence profile at the inlet of the domain is provided by a precursor simulation carried out with identical conditions in the absence of the farm. After the precursor simulation reaches a fully developed turbulence regime, a region of length $L_\mathrm{fr}$ is duplicated from the precursor simulation on the fringe region of the actual simulation at the end of every time step. Then, any variable $\phi$ (i.e. velocity and potential temperature) in the fringe region is determined as a weighted average of fields in the precursor and actual simulations \citep[also see][]{Stevens2014RE},
\begin{equation} \label{eq:weight_avg}
    \phi(x,y,z,t) = f(x) \cdot \phi_\mathrm{pre}(x,y,z,t) + \left[1-f(x)\right] \cdot \phi_\mathrm{act}(x,y,z,t), 
\end{equation}
in which $\phi_\mathrm{pre}$ and $\phi_\mathrm{act}$ are, respectively, the field in the precursor and actual domains, and $f(x)$ is the weighting function expressed as,
\begin{equation} \label{eq:weight_func}
  f(x) = \left\{
    \begin{array}{ll}
      \frac{1}{2}\left[ 1 - \cos{\left(\upi \frac{x-x_s}{x_e-x_s} \right)} \right], 
      & x_s \le x \le x_e \\[2pt]
      1,         & x > x_e.
    \end{array} \right.
\end{equation}
Here, $x$ represents the downstream position, $x_s=L_x-L_u-L_\mathrm{fr}$ is the starting point of the fringe region, $x_e = L_x-L_u-\frac{1}{4}L_\mathrm{fr}$ is the position beyond which $\phi=\phi_\mathrm{pre}$. The length of the fringe region must be large enough to enable a smooth transition of the field $\phi$ from the farm wake flow to the inflow condition. To avoid any possible upstream influence from the fringe region, only solutions up to $x=x_s-3h_\mathrm{b}$ are analyzed.


\subsection{Simulation parameters} \label{sec:simulation}
Our major goal is to report new flow features that develop around suspended aquafarms under realistic oceanic conditions. Therefore, instead of exploring the vast parameter space of possible ocean states (e.g. varying degrees of wind, waves, currents, and surface buoyancy forcing, etc.), we only focus on one set of very typical conditions encountered in the deep ocean. The flow is driven by two main forcings, i.e. the overlying atmospheric flow and a geostrophic current, in a uniformly rotating environment with the Coriolis frequency $f=1.0 \times 10^{-4}\ \mathrm{s^{-1}}$ (corresponding to a latitude of 45$^{\circ}$N). The simulation parameters are chosen to be the same as those used in \citet{McWiliams:1997jfm}, which serves as benchmark case in the literature on Langmuir turbulence \citep{Polton2008GRL, Skitka2020JPO}.  A constant wind stress $\tau_w=0.37\ \mathrm{N\ m^{-2}}$ is applied at the air-sea interface and aligned with the wave field in the downstream direction. The corresponding wind speed at 10-m height is $U_{10}=5\ \mathrm{m\ s^{-1}}$, and the friction velocity at the ocean surface is $u_*=6.1\times10^{-3}\ \mathrm{m\ s^{-1}}$. The wave field consists of monochromatic waves with wavelength $\lambda=60\ \mathrm{m}$ (corresponding to a wave period $T_w=6.2\ \mathrm{s}$) and amplitude $a_w=0.8\ \mathrm{m}$, corresponding to $U_s=0.068\ \mathrm{m\ s^{-1}}$. The resulting turbulent Langmuir number $La_t=\sqrt{u_*/U_s}=0.3$, which is typical for wind-wave equilibrium conditions in the open ocean \citep{Belcher2012grl}.

A geostrophic current $u_g=0.2\ \mathrm{m\ s^{-1}}$ in the downstream direction is superimposed on the flow field to represent the effect of mesoscale flow features, which are considered to behave as a constant flow on the time and spatial scales of interest here (5 hrs and a few kilometers). The upper mixed layer is bounded by a stably stratified layer below with a constant temperature gradient $\mathrm{d}\theta/\mathrm{d}z = 0.01\ \mathrm{K\ m^{-1}}$. Since surface heating or cooling would add another layer of complexity associated with buoyancy effects on turbulence, we assume zero buoyancy flux at the ocean surface for the simulations considered here.

Table \ref{tab:1} summarizes the simulation parameters and resolution of six different cases considered here. In the table, $N_x$, $N_y$, and $N_z$ are the number of grid points in the $x$, $y$, and $z$ directions, respectively. Simulation cases CLT/LT and CST/ST represent the modelling of Langmuir turbulence and pure shear-driven turbulence in the presence/absence of macroalgal farm, respectively. These four cases are carried out to evaluate the effects of macroalgae canopy and the role of surface gravity waves on the flow features. The shear-driven cases CST and ST are conducted in the absence of any surface wave forcing, i.e. the wave-induced Stokes drift velocity is zero. For a boundary layer flow within and under a suspended canopy of finite size, whether or not the boundary layer can reach a fully developed stage depends on the length of the canopy \citep{Tseung:2016efm}. Thus, Langmuir turbulence in the presence of a longer farm ($L_\mathrm{MF}=800$ m), referred to as case CLTL, is performed to explore the limit of fully developed flow. We focus mostly on the results of the CLT simulation and use CLTL only when investigating the downstream flow development. The mesh is uniformly distributed, with a horizontal resolution $\Delta_h=2$ m and vertical resolution of $\Delta_z=0.5$ m. To confirm that the resolution is sufficient, case CLTF is performed under the same setup as CLT, but with finer-scale resolution (twice the resolution) in all three directions.

Cases LT and ST are initialized with a converged solution based on a initial mixed layer depth (MLD) of 20 m, from which the inertial oscillations have been removed. The turbulence is confined to the upper mixed layer and the water column below is stably stratified. Then, cases LT and ST serve as precursor simulations to provide time-varying turbulent inflow conditions for cases CLT(F/L) and CST, respectively. The simulations CLT(F/L) and CST are first carried out for 15,000 s to allow for the adjustment of the surface boundary layer to the macroalgae canopy. After the turbulent flow has reached a quasi-equilibrium state, the flow field is averaged over another 9,000 s to obtain turbulence statistics. Finally we note that even though turbulence scales with $u_*/La_T^{2/3}$ in Langmuir turbulence \citep{Grant2009JPO}, we use the surface friction velocity $u_{*}$ as the scaling velocity throughout the paper to facilitate the comparison between Langmuir and shear-driven turbulence. 

\begin{table}
  \begin{center}
\def~{\hphantom{0}}
  \begin{tabular}{lcccccc}
       Case  & Canopy & Wave & $La_t$  & $L_\mathrm{MF} (\mathrm{m})$ &
       $L_x (\mathrm{m}) \times L_y (\mathrm{m}) \times L_z (\mathrm{m})$ & $N_x \times N_y \times N_z$ \\[3pt]
       CLT  & Yes & Yes & 0.3 & 400 & 800 $\times$ 208 $\times$ 120 & 400 $\times$ 104 $\times$ 240 \\
       CLTF & Yes & Yes & 0.3 & 400 & 800 $\times$ 208 $\times$ 120 & 800 $\times$ 208 $\times$ 480 \\
       CLTL & Yes & Yes & 0.3 & 800 & 1200 $\times$ 208 $\times$ 120 & 600 $\times$ 104 $\times$ 240 \\
       CST  & Yes & No  & N/A & 400 & 800 $\times$ 208 $\times$ 120 & 400 $\times$ 104 $\times$ 240 \\
       LT   & No & Yes & 0.3  & N/A & 400 $\times$ 208 $\times$ 120 & 200 $\times$ 104 $\times$ 240 \\
       ST   & No & No & N/A & N/A & 400 $\times$ 208 $\times$ 120 & 200 $\times$ 104 $\times$ 240\\
  \end{tabular}
  \caption{Parameters of the LES runs}
  \label{tab:1}
  \end{center}
\end{table}

\begin{figure}
    \centerline{\includegraphics[width=1.0\linewidth]{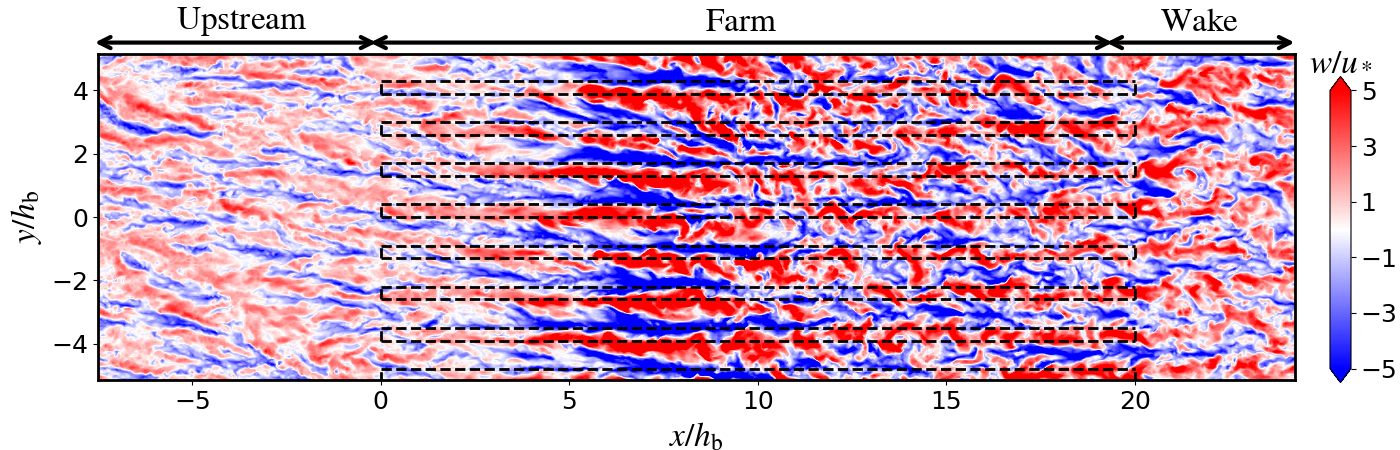}}
    \caption{Snapshot of the normalized vertical velocity $w/u_*$ on a horizontal plane ($z=-0.25h_\mathrm{b}$) for case CLTF. The black dashed rectangles represent the region occupied by macroalgae canopy. The blue and red color indicate downwelling and upwelling regions.}
\label{fig:snapshot}
\end{figure}

A snapshot of the vertical velocity $w/u_*$ on a horizontal plane at $z=-0.25h_\mathrm{b}$ for case CLTF is shown in figure \ref{fig:snapshot}. The elongated streaks of downward vertical velocity readily observed upstream from the farm leading edge are signatures of Langmuir circulations. They are oriented to the right of the wind direction (i.e. $x$-direction), and are transient structures that are continuously generated and dissipated. As the OML flows into the farm, however, a persistent pattern with stronger downward and upward velocities alternating laterally is clearly seen, roughly parallel to the canopy rows. The magnitude of $w/u_*$ within the farm region can be as large as 8.0 (the colorbar has been saturated), while the typical values for Langmuir and shear turbulence in the absence of the farm for the same ocean conditions are 1.6 and 0.75, respectively \citep[e.g. see][]{McWiliams:1997jfm}. This quasi-stationary pattern of alternating upwelling and downwelling regions indicates the existence of counter-rotating cells, hereafter referred to as attached Langmuir circulations (as discussed below). These secondary flow structures extend beyond the trailing edge in the farm wake zone.

\subsection{Flow decomposition} \label{sec:decomposition}
The statistics for cases LT and ST are obtained by averaging both temporally and horizontally, indicated by $\left \langle \overline{\ \cdot\ }\right \rangle$. Note that the time average and spatial average are indicated by an overbar and a pair of angled brackets, respectively. The physical quantities for CLT and CST are first averaged in the temporal dimension. Because of the three-dimensional spatial heterogeneity of the flow, these time-averaged statistics are subject to larger random errors than the spatial-temporal averaging used for cases LT and ST. Thus, either a spatial or phase averaging operation in the cross-stream $y$ direction is also used, indicated respectively by $\left \langle\ \right \rangle_y$ or $\left \langle\ \right \rangle_p$. Given the idealized cross-stream canopy heterogeneity, the cross-phase average defined here, different from the wave-phase average introduced in deriving equation \eqref{eq:momentum}, corresponds to averaging over equivalent positions in cross-stream phases. For any time-averaged field $\overline{\phi}$, the cross-phase averaging can be expressed as,
\begin{equation} \label{eq:phaseAvg}
    \left \langle{\overline{\phi}}\right \rangle_p (x,y,z) =\frac{1}{N} \sum_{n=0}^{N-1} {\overline{\phi}}(x,y+nS_\mathrm{MF},z), 
\end{equation}
where $N=8$ is the number of macroalgae rows. 

Hereafter, we use the cross-stream average to define the (primary) mean field $\langle \overline{\phi} \rangle_{y}(x,z)$, and the deviations from the mean field are decomposed into a secondary-flow component and a transient component. Thus, instantaneous flow quantities, such as the velocity field $\boldsymbol{u}$, can be represented by,
\begin{equation} \label{eq:decomposition}
    \boldsymbol{u}=\overline{\boldsymbol{u}}+\boldsymbol{u}'=\left<\overline{\boldsymbol{u}}\right>_y+\overline{\boldsymbol{u}}^c+\boldsymbol{u}', 
\end{equation}
Here, $\boldsymbol{u}'$ denotes the transient fluctuation from $\overline{\boldsymbol{u}}$, while the secondary-flow disturbance $\overline{\boldsymbol{u}}^c=\overline{\boldsymbol{u}}-\left<\overline{\boldsymbol{u}}\right>_y$ is stationary in time and represents the lateral structure of the time-averaged velocity field induced by the farm geometry. As the transient fluctuation and secondary-flow disturbance are uncorrelated, the covariance between the velocity component $u_i$ and any field $\phi$ can be written as,
\begin{equation} \label{eq:flux_decomposition}
    \left<\overline{u_i\phi}\right>_y=\left<\overline{u_i}\right>_y\left<\overline{\phi}\right>_y+\left<\overline{u_i}^c \overline{\phi}^c\right>_y+\left<\overline{u_i'\phi'}\right>_y,
\end{equation}
The three terms on the RHS represent the of contributions from the mean flow, the secondary-flow part, and the transient part, respectively.

Finally, in some cases we further average results in the vertical direction (depth-averaged), from the free surface $z = 0$ to a fixed depth $z = z_t$ with $z_t=-2h_b$, which are then represented by 
\begin{equation} \label{eq:z_avg}
  \langle \overline{\boldsymbol{u}} \rangle_{yz} = \frac{1}{|z_t|}\int_{z_t}^{0} \langle \overline{\boldsymbol{u}} \rangle_y \mathrm{d}z.
\end{equation}

\section{Langmuir turbulence in the presence of canopy} \label{sec:3}

\subsection{Adjustment of the mean flow} \label{SS:wmean}

\begin{figure}
  \centering\includegraphics[width=1.0\linewidth]{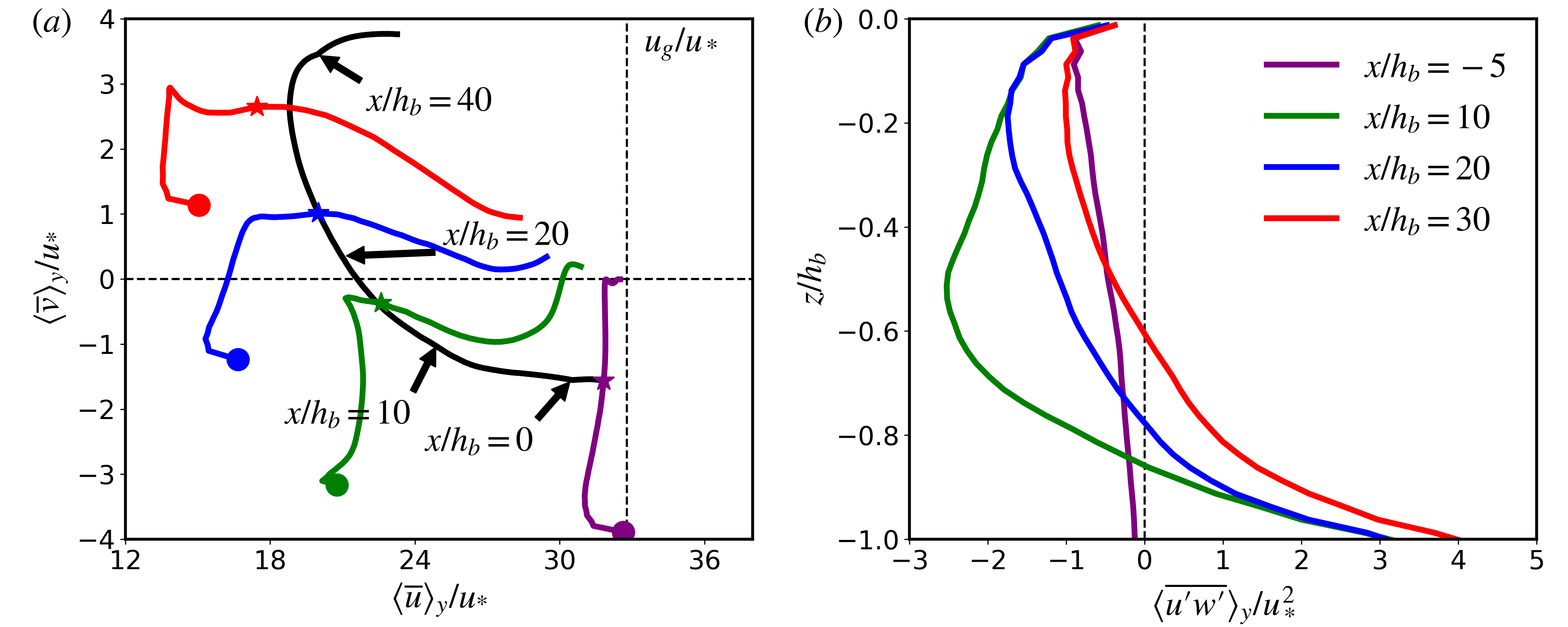}
  \caption{(\textit{a}) Hodographs of the mean velocity vector ($\langle \overline{u} \rangle_y,\ \langle \overline{v} \rangle_y$) in the vertical at four different downstream positions as noted in the legend are also included, and downstream variation of the depth-averaged mean velocity vector ($\langle \overline{u} \rangle_{yz},\ \langle \overline{v} \rangle_{yz}$) (black line); (\textit{b}) profiles of the resolved momentum stress $\left<\overline{u'w'}\right>_y$ at these selected downstream locations. Circles indicate values at the surface $z/h_b = 0$, and asterisks indicate the canopy bottom $z/h_b = -1$.}
\label{fig:mean_advec}
\end{figure}

The OML undergoes significant changes as it approaches and flows over the farm. Here, we present the mean flow for case CLTL to offer a more complete picture of the spatial development of the upper OML. Figure \ref{fig:mean_advec}\textit{a} shows the hodographs of the mean horizontal velocity vector ($\langle \overline{u} \rangle_y,\ \langle \overline{v} \rangle_y$) at four different downstream positions. Upstream from the canopy leading edge ($x/h_b = -5$, purple line), the hodograph follows a typical Stokes-Ekman spiral in Langmuir turbulence, with the cross-stream velocity pointing to the right of the wind stress (i.e., $\langle \overline{v} \rangle_y<0$) and most of the shear located near the surface (the horizontal velocity is nearly uniform within most of the OML depth due to strong vertical mixing). As the flow moves into the farm ($x/h_b = 10,\ , 20\ , 30$), the hodographs become very distorted due to the large effect of the canopy drag. Despite the very complex behavior of the mean flow, some features are noteworthy. At $x/h_b = 20$ (blue line), the cross-stream component of the flow switches direction within the OML, and at $x/h_b = 30$ (red line), the cross-stream flow is completely reversed (i.e., to the left of the wind direction within the entire depth of the OML). Also included in the figure is the downstream variation of the depth-averaged horizontal velocity vector ($\langle \overline{u} \rangle_{yz},\ \langle \overline{v} \rangle_{yz}$) (black line). Downstream from the leading edge, we can see that the depth-averaged mean flow direction changes sign at $x/h_b \approx 18$, indicating a change in the direction of cross-stream advection within the farm.

The overall change in the direction of the cross-stream flow can be understood based on the differences of surface and bottom boundary layers in the presence of the rotation. In the northern hemisphere, the horizontal transport is oriented to the right of the wind stress in surface Ekman layers, and to the left of the main current in bottom Ekman layers \citep{McWilliams2006}. In the present case, the canopy introduces a vertically distributed drag that is more pronounced near the bottom of the farm (where the LAD and the mean velocities are larger). Therefore, the sign of $\langle \overline{v} \rangle_y$ depends critically on the relative importance of shear stresses at the top and bottom of the farm. Specifically, if the stress near the ocean surface dominates over the stress around the canopy bottom, then $\langle \overline{v} \rangle_y$ is aligned to the right of the wind as in the wind-stress driven mixed layer \citep{McWiliams:1997jfm}; if the stress at the canopy bottom prevails, then the flow behaves more like a bottom boundary layer above the canopy bottom and $\langle \overline{v} \rangle_y$ is directed to the left of the wind (right of the bottom stress) \citep{Taylor2008b}. Because the former scales with $u_*$ and the latter with $u_g$, we expect the flow behavior for a fixed canopy configuration to depend on the ratio $u_g/u_*$. Figure \ref{fig:mean_advec}\textit{b} shows the vertical profiles of $\left<\overline{u'w'}\right>_y$ at the selected four downstream locations. It clearly shows that the turbulence within the farm has not reached a fully developed state in the downstream direction, and the complexity of hodographs from figure \ref{fig:mean_advec}\textit{a} also reflects this fact. Along the $x-$direction, the flow transitions from a surface-stress-dominated regime to a bottom-stress-dominated flow, which explains the switch in mean cross-stream flow direction shown in figure \ref{fig:mean_advec}\textit{a}.

\begin{figure}
    \centerline{\includegraphics[width=1.0\linewidth]{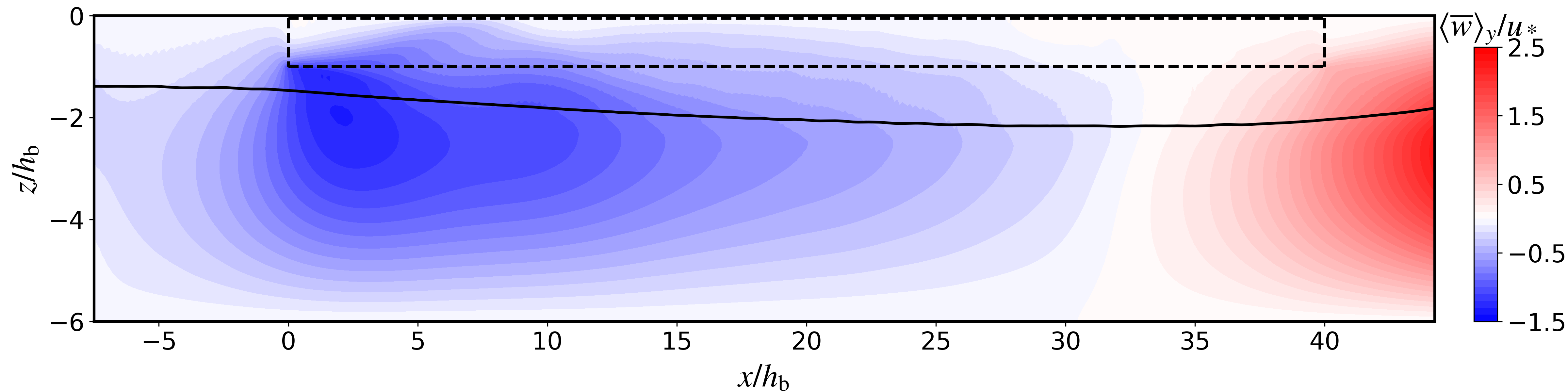}}
    \caption{The time- and cross-stream-averaged vertical velocity $\langle \overline{w} \rangle_y$, nomalized by $u_{*}$, for case CLTL along the $x-z$ plane. The black dashed rectangle represents the location where the macroalgae is planted. The black solid line marks the mixed layer depth, which is defined as the position where the temperature exceeds a certain percentage of the mixed layer value.} 
\label{fig:w_mean}
\end{figure}

Figure \ref{fig:w_mean} displays the mean vertical velocity $\langle \overline{w} \rangle_y/u_{*}$ along the $x-z$ plane for case CLTL. The region occupied by the macroalgae canopy is highlighted in a dashed rectangle. The $\langle \overline{w} \rangle_y/u_{*}$ exhibits a small value near the inlet, which implies that the macroalgae farm poses a minor impact on the inflow. As the flow approaches the macroalgae farm, the canopy drag obstructs the fluid. The associated pressure gradient across the leading edge decelerates the flow within a region upstream of the canopy \citep[termed the ``impact region" in][]{Belcher2003jfm} and induces a downward vertical motion under the canopy near the leading edge by continuity. Similarly, the pressure drop across the trailing edge causes the wake flow return to its inflow profile, leading to an upward motion into the wake of the farm. 

The solid line in figure \ref{fig:w_mean} illustrates the downstream variation of the mixed layer depth (MLD), denoted as $z_i$. As it develops downstream, the shear turbulence near the bottom of the macroalgae canopy gradually erodes the stratification by entraining denser water into the upper mixed layer. Here, we define MLD as the location at which the potential temperature first exceeds a certain percentage of the mixed layer temperature $\theta_\mathrm{ML}$. Thus
\begin{equation}
  z_i = \{z: \langle \theta \rangle_y (x, z) - \theta_\mathrm{ML} = \chi \theta_\mathrm{ML}\}
\end{equation}
where $\chi$ is a predefined constant. This definition is adapted from the potential temperature contour method in \citet{Sullivan:1998jas}. The downstream evolution of the MLD indicates that, for the present configuration in which the MLD is comparable to the depth of the backbone line, the shear layer at the bottom of the farm creates a local perturbation in the depth of the OML, which seems to recover downstream from the farm.

\subsection{Attached Langmuir circulations} \label{sec:standing}

\begin{figure}
    \centerline{\includegraphics[width=1.0\linewidth]{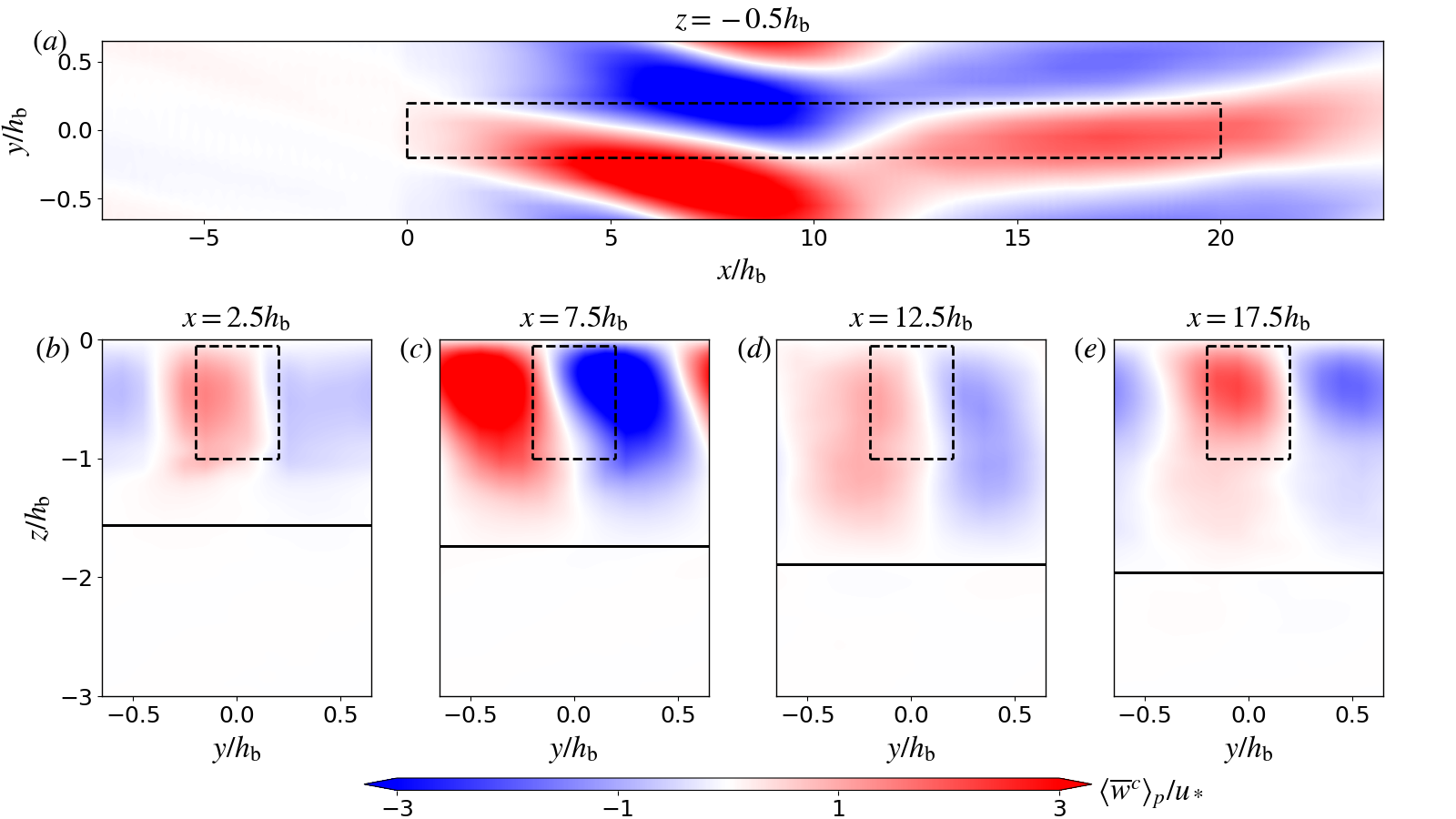}}
    \caption{The normalized secondary-flow part of vertical velocity $\langle \overline{w}^c \rangle_p/u_{*}$, averaged over time and cross-phase, for case CLT on a $x-y$ plane at $z=0.5h_\mathrm{b}$ (\textit{a}), and $y-z$ plane (facing upstream) at $x=2.5h_\mathrm{b}$ (\textit{b}), $x=7.5h_\mathrm{b}$ (\textit{c}), $x=12.5h_\mathrm{b}$ (\textit{d}), and $x=17.5h_\mathrm{b}$ (\textit{e}). The black solid line marks the mixed layer depth. The black dashed rectangles represent the location where the macroalgae is planted. The velocity has been cross-phase-averaged and remapped to the entire plane. The extreme colors of the colorbar are saturated to highlight the spatial variation of the strength of the cell pattern.}
\label{fig:w_se}
\end{figure}

Figure \ref{fig:w_se} shows the contours of the secondary-flow part of the vertical velocity $\langle \overline{w}^c \rangle_p/u_{*}$ for case CLT in the cross-sections noted in the caption. In the figure, we can observe a regular pattern of $\langle \overline{w}^c \rangle_p$ alternating between positive and negative values along the cross-stream direction, indicating the steady upwelling and downwelling motions induced by the presence of the canopy. This organized pattern is the signature of pairs of steady counter-rotating circulatory flows with axis approximately aligned in the streamwise direction. These upweling and downweling regions extend to the bottom of the OML. We infer that these flows are primarily driven by the wave-current interaction since these features are not observed in the shear-driven case CST (not shown). We refer to these flow structures as attached Langmuir circulations because (i) their position is determined by the spatial structure of the canopy, and (ii) their formation depends critically on the wave-induced Stokes drift via a mechanism that resembles the CL type 2 instability, which will be described in section \ref{sec:mechanism}.

While the standard Langmuir circulations appear as unsteady structures that move around in the flow (see figure \ref{fig:snapshot}), the attached Langmuir cells are more steady and regularly spaced. For the present canopy configuration, the separation between neighboring pairs of attached Langmuir cells is determined by the lateral spacing between consecutive rows of macroalgae elements, but test runs suggest that this could change if the distance between canopy rows is significantly larger (not shown). As the flow moves downstream, the strength of the canopy-induced Langmuir circulations exhibits a non-monotonic variation. The downwelling velocity reaches its maximum value at $x \approx 7.5h_\mathrm{b}$ with a magnitude of approximately $8u_*$ (figure \ref{fig:w_se}\textit{b}). The cell pattern then gradually decays until $x \approx 12.5h_\mathrm{b}$ (figure \ref{fig:w_se}\textit{c}), and recovers at a lower level further downstream towards the trailing edge of the farm (figure \ref{fig:w_se}d). The orientation of Langmuir cells can be identified by the elongated downwelling streaks. Owing to the non-zero component in the mean cross-stream velocity (see figure \ref{fig:mean_advec}\textit{a}), the canopy-attached Langmuir circulations are oblique to the downstream direction. The upwelling and downwelling bands are mildly deflected to the right of the wind for $x/h_\mathrm{b}< 10.0$, and then aligned somewhat to the left of the wind for $x/h>10.0$, in agreement with the change in cross-stream velocity discussed in the previous section. This complex pattern is discussed further in section \ref{sec:mechanism}, where results for the long farm case (case CLTL) are presented. 

\begin{figure}
    \centerline{\includegraphics[width=1.0\linewidth]{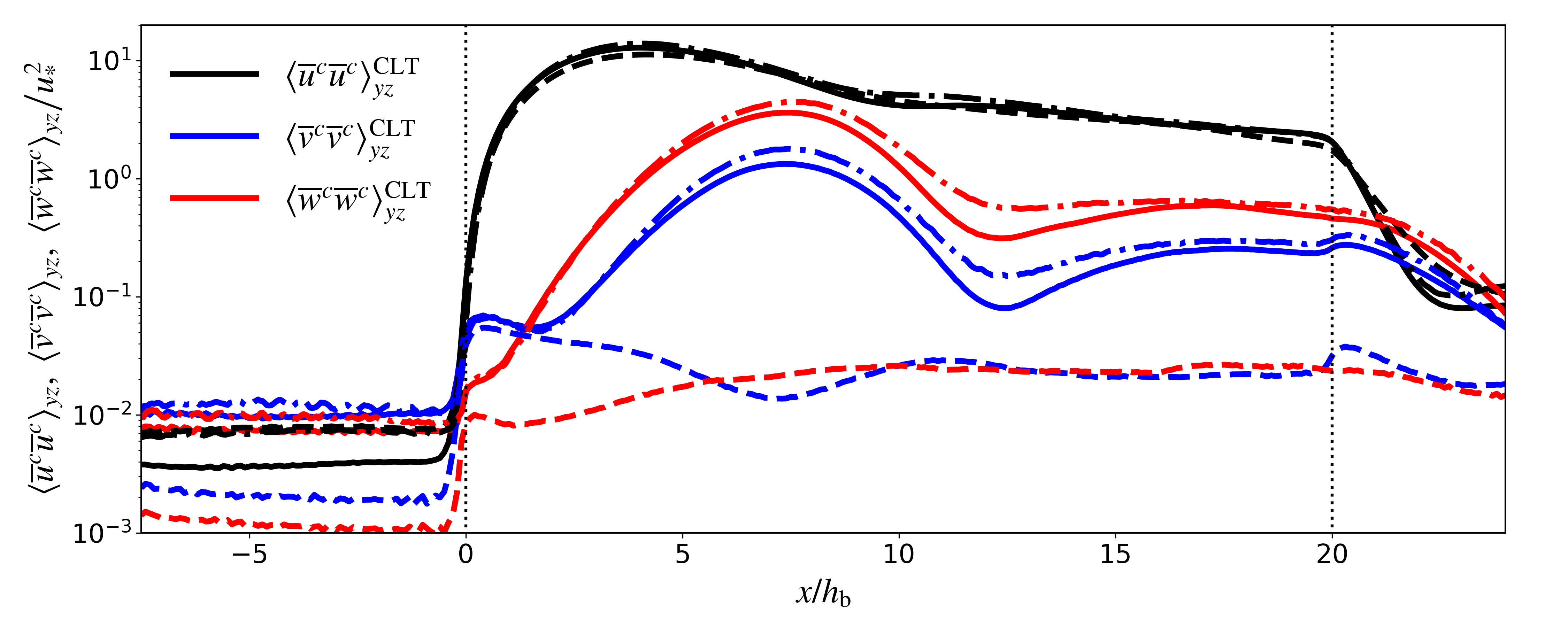}}
    \caption{The cross-stream- and depth-averaged secondary-flow part of velocity variances for CLT (solid lines) and CST (dashed lines), together with the results from CLTF (dash-dotted line). The vertical dotted lines mark the leading and trailing edge of the farm.}
\label{fig:var_se}
\end{figure}

As clearly seen in figures \ref{fig:snapshot}-\ref{fig:w_se}, the flow field has not reached a fully developed state at the trailing edge of the farm (true for both the short and long farms). For canopy flows, the canopy-drag length is defined as $L_c = \left (\frac{1}{2} C_D \overline{a} \right )^{-1}$ where $\overline{a} = W_\mathrm{MF}/(S_\mathrm{MF}h_\mathrm{b})\int_{-h_\mathrm{b}}^0 a(z)P_x\ \mathrm{d} z$ is the effective FAD. This length scale neglects the vertical and horizontal structure of the canopy, and characterizes the distance over which the flow adjusts to the mean drag of canopy elements \citep{Belcher2003jfm, Rominger2011JFM}. The values reported in section \ref{sec:2} yield $\overline{a}h_b \approx 7.0$ and $L_c \approx 19.2h_\mathrm{b}$. Note that the short and long farms have lengths of approximately equal to $L_c$ and $2L_c$, suggesting that the upper mixed layer flow does not fully adjust to the canopy in these two cases. 

To quantify the strength of the attached Langmuir circulations, we focus on the three components of velocity variances due to the contribution from the secondary flow. Figure \ref{fig:var_se} shows the downstream variation of the depth-averaged mean velocity variances for cases CLT and CST. The results from CLTF are also included to examine the sensitivity to grid resolution. The comparison shows that the finer resolution simulation (CLTF) yields relatively larger variances than CLT in all three velocity components, but the overall variations observed in CLTF conform qualitatively to those in CLT. Thus, we consider the simulations with moderate resolution (CLT and CST, etc.) to be a good starting point to explore Langmuir turbulence in the presence of marine plants. It is interesting to note that $\langle \overline{u}^c \overline{u}^c \rangle_{yz}$ shows negligible differences within the farm between CLT and CST, suggesting that the canopy effect on the streamwise velocity component of the secondary flow is not impacted by the surface waves. This also indicates that $\langle \overline{u}^c \overline{u}^c \rangle_{yz}$ is dominated by the lateral variation in mean velocity due to the spatially varying drag. For case CST, the magnitudes of $\langle \overline{v}^c \overline{v}^c \rangle_{yz}$ and $\langle \overline{w}^c \overline{w}^c \rangle_{yz}$ within the canopy exceed their upstream levels by roughly an order of magnitude, suggesting that the presence of canopy rows leads to some secondary circulations driven by adjustment to the canopy drag, which may also be impacted by spatial variation in the turbulent stresses (i.e. Prandtl's secondary flow of the second kind) \citep{Bradshaw1987ARFM}. In the simulation with the Stokes drift (case CLT), however, $\langle \overline{v}^c \overline{v}^c \rangle_{yz}$ and $\langle \overline{w}^c \overline{w}^c \rangle_{yz}$ are about two orders of magnitude greater than that in the Stokesless simulation (case CST). The downstream enhancement and reduction of $\langle \overline{v}^c \overline{v}^c \rangle_{yz}$ and $\langle \overline{w}^c \overline{w}^c \rangle_{yz}$ within the canopy for case CLT are consistent with the pattern of the vertical velocity in figure \ref{fig:w_se}. Therefore, we conclude that, for the present configuration, the presence of Stokes drift is a key factor enabling the mean streamwise flow structure induced by the farm drag to develop into strong secondary circulations. As discussed above, these eddies are roughly two-dimensional with centerline approximately aligned in the downstream direction, justifying the nomenclature ``attached Langmuir circulations". Based on these results, hereafter we interpret the streamwise component of the secondary flow as a product of the spatial structure of the canopy drag, and the crosswise and vertical components of the secondary flow in simulations with Stokes drift as attached Langmuir circulations.

\subsection{Langmuir turbulence intensity} \label{sec:intensity}

\begin{figure}
    \centerline{\includegraphics[width=1.0\linewidth]{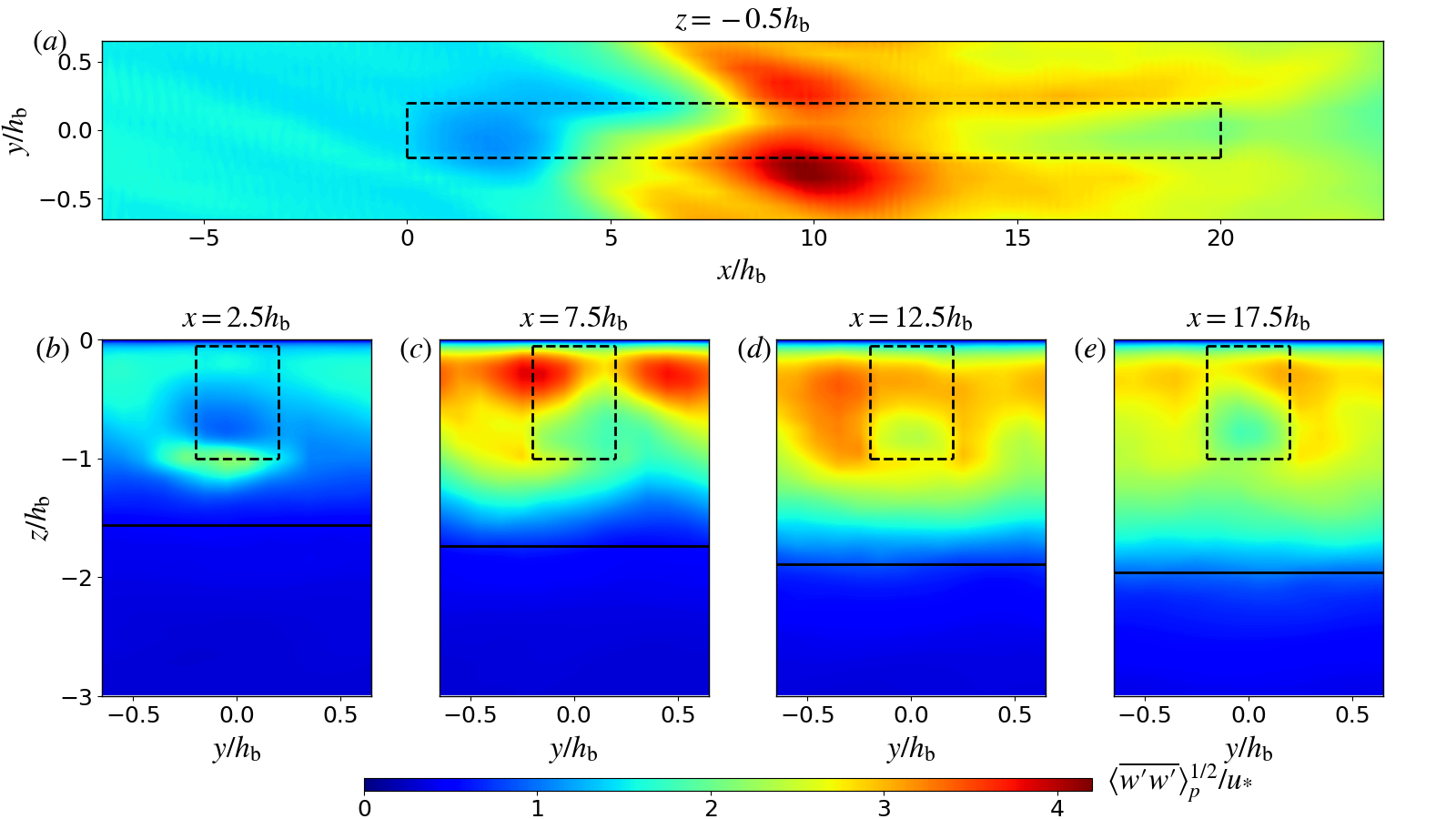}}
    \caption{The transient part of the vertical velocity variance $\langle \overline{w'w'} \rangle^{1/2}_p/u_{*}$ for case CLT in the $x-y$ plane at $z=0.5h_\mathrm{b}$ (\textit{a}), and $y-z$ plane at $x=2.5h_\mathrm{b}$ (\textit{b}), $x=7.5h_\mathrm{b}$ (\textit{c}), $x=12.5h_\mathrm{b}$ (\textit{d}), and $x=17.5h_\mathrm{b}$ (\textit{e}).}
\label{fig:wrms_CLT}
\end{figure}

Langmuir turbulence intensity is often characterized by large vertical velocity variance. Our interest is centered on how the macroalgae farm alters the spatial evolution of turbulence levels and associated turbulent mixing efficiency. In the figure \ref{fig:wrms_CLT}, we plot the time- and cross-phase-averaged vertical velocity variance due to transient eddies $\langle \overline{w'w'} \rangle^{1/2}_p/u_{*}$ for case CLT. Similar to that in standard Langmuir turbulence, the vertical intensity $\langle \overline{w'w'} \rangle^{1/2}_p/u_{*}$ peaks at a subsurface level, even in the presence of a shear layer near the surface due to canopy discontinuity (top 1m). In the nearfield downstream from the leading edge ($0<x/h_\mathrm{b}<4$), $\langle \overline{w'w'} \rangle^{1/2}_p/u_{*}$ is decreased within the canopy and increased near the canopy bottom (figure \ref{fig:wrms_CLT}a and b). This is because the canopy drag dampens the vertical kinetic energy within the canopy, but the shear layer at the canopy bottom can inject additional energy from the mean flow via shear production (see section \ref{sec:energetics}). Further downstream, $\langle \overline{w'w'} \rangle^{1/2}_p/u_{*}$ first increases, with the maximum value occurring at $9<x/h_\mathrm{b}<11$ (figure \ref{fig:wrms_CLT}a), and then decreases towards the trailing edge. The energetics of the upper mixed layer, which will be covered in section \ref{sec:energetics}, suggest that the enhancement and reduction of $\langle \overline{w'w'} \rangle^{1/2}_p/u_{*}$ are mainly determined by two processes: (i) the energy exchanges with the attached Langmuir circulations and (ii) the shear production associated with the lateral/vertical shear in streamwise velocity caused by the canopy structure. In the downstream cross-section (figure \ref{fig:wrms_CLT}$\textit{c-d}$), a clear pattern emerges with increased $\langle \overline{w'w'} \rangle^{1/2}_p/u_{*}$ at the bottom and outer edge of the canopy rows and reduced intensity in the lower half of the canopy row where the leaf area density is high (figure \ref{fig:morphology}b).

\begin{figure}
    \centerline{\includegraphics[width=1.0\linewidth]{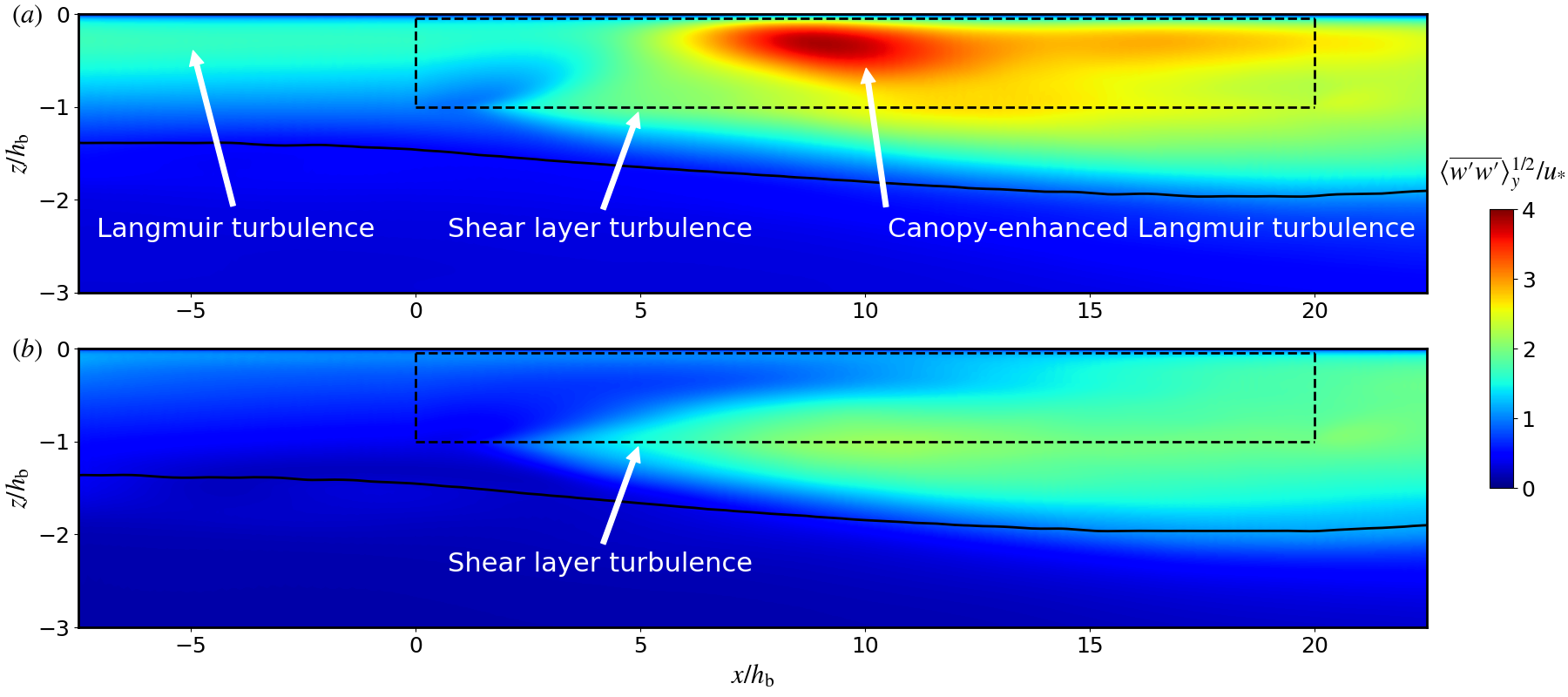}} 
    \caption{The transient part of the vertical velocity standard deviation $\langle \overline{w'w'} \rangle^{1/2}_y/u_{*}$ for CLT (\textit{a}) and CST (\textit{b}) in the $x-z$ plane. The black solid line marks the mixed layer depth.}
\label{fig:wrms_comp}
\end{figure}

Figure \ref{fig:wrms_comp} shows the comparison of the RMS of the transient vertical velocity fluctuation $\langle \overline{w'w'} \rangle^{1/2}_y/u_{*}$ between Langmuir (case CLT, upper panel) and shear turbulence (case CST, lower panel). Upstream from the leading edge, $\langle \overline{w'w'} \rangle^{1/2}_y$ from case CLT is about twice as large as that from case CST. This is because Langmuir turbulence yields significantly higher vertical velocity intensity compared to the pure shear-driven turbulence scenario \citep{McWiliams:1997jfm, D'Asaro2001JPO, Harcourt2008JPO}. In the absence of surface wave forcing (figure \ref{fig:wrms_comp}b), the contour of $\langle \overline{w'w'} \rangle^{1/2}_y/u_{*}$ is similar to what is expected for open-channel flow over a suspended canopy \citep[see figure 16e in][]{Tseung:2016efm}. The shear layer at the canopy bottom grows continually downstream and penetrates upward into the canopy, leading to the augmentation of $\langle \overline{w'w'} \rangle^{1/2}_y$ within the growing shear layer. Towards the end of the farm, the shear layer penetrates over the entire canopy depth, a phenomenon that usually occurs for sparse canopies \citep{Nepf:2012ARFM}. Interestingly, in the simulation that includes the wave-induced Stokes drift  (figure \ref{fig:wrms_comp}a), the shear layer turbulence seems to merge with Langmuir turbulence at around $x/h_\mathrm{b} \approx 4$, and the turbulence levels near the ocean surface are further enhanced within the canopy (for $6<x/h_\mathrm{b}<12$) as compared to the Stokesless counterpart (figure \ref{fig:wrms_comp}b). This can be attributed to the presence of attached Langmuir circulations described above in section \ref{sec:standing}. This difference between the two cases also confirms that the enhancement of $\langle \overline{w'w'} \rangle^{1/2}_p/u_{*}$ in figure \ref{fig:wrms_CLT} is due to the turbulence modulation by the attached Langmuir circulations.

\begin{figure}
    \centerline{\includegraphics[width=1.0\linewidth]{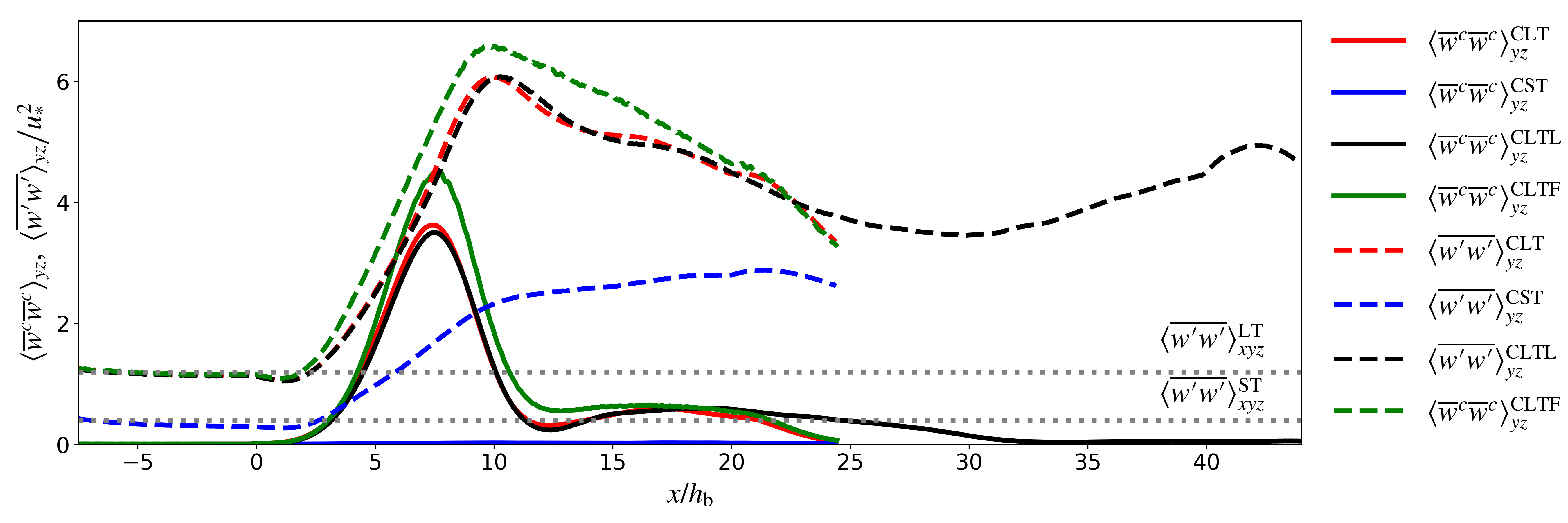}} 
    \caption{Downstream variations of $\langle \overline{w'w'} \rangle_{yz}/u_{*}^2$ (dashed line) and $\langle {\overline{w}^c \overline{w}^c} \rangle_{yz}/u_{*}^2$ (solid line) for CLT (red), CST (blue), CLTF (green), and CLTL (black). The end of the farm is located at $x/h_\mathrm{b}=20$ for CLT/CST/CLTF, and $x/h_\mathrm{b}=40$ for CLTL. The grey dotted lines mark the values of the depth-averaged vertical velocity variance for normal Langmuir turbulence (case LT, upper line) and pure shear-driven turbulence (case ST, lower line).}
\label{fig:wrms_line_comp}
\end{figure}

Since transient eddies and attached Langmuir circulations coexist as the fluid impinges upon and flows over the farm (figure \ref{fig:snapshot}), it is desirable to compare the energy associated with transient eddies from that of attached Langmuir circulations. In figure \ref{fig:wrms_line_comp}, we plot the vertical velocity variances due to the contribution from the transient eddies and attached Langmuir circulations as noted in the caption. Again, only some minor differences exist between CLT and CLTF within the farm region, building confidence on the use of the coarser simulations to analyze the flow. To evaluate if the flow has fully adjusted to the canopy towards the end of the farm in cases CLT, the results from CLTL are also shown. The discrepancies between cases CLT and CLTL (black and red lines) are mainly located near the end of the farm in CLT ($x/h_\mathrm{b} = 20$) due to the trailing edge effect. As the farm extends further downstream (case CLTL, $L_\mathrm{MF}=40h_\mathrm{b}$), $\langle \overline{w}^c \overline{w}^c \rangle^\mathrm{CLT}_{yz}$ does not become uniform but still evolves in the streamwise direction within the farm (black solid line). It is observed that the attached Langmuir circulations gradually attenuate in strength from $x/h_\mathrm{b} \approx 20$ and eventually fade away at $x/h_\mathrm{b} \approx 32$ (black solid line). This suggests that their existence is a result of flow adjustment to the suspended farm of finite size rather than a fully developed state. While the attached Langmuir circulations disappear, the vertical velocity variance of transient eddies for case CLTL $\langle \overline{w'w'} \rangle^\mathrm{CLTL}_{yz}$ is increasing from $x/h_\mathrm{b} \approx 30$ towards the end of the farm (black dashed line). The enhanced $\langle \overline{w'w'} \rangle^\mathrm{CLTL}_{yz}$ of transient eddies is mainly attributed to the canopy shear in the horizontal direction, which no longer assists the generation of attached Langmuir circulations as the flow has reached an equilibrium state. Except in the nearfield downstream of the leading edge, $\langle \overline{w'w'} \rangle^\mathrm{CLT}_{yz}$ is much larger than $\langle \overline{w}^c \overline{w}^c \rangle^\mathrm{CLT}_{yz}$ throughout the remaining part of the farm.

\subsection{Comparison against standard Langmuir circulations}\label{sec:condition}

To compare the attached Langmuir cells with the traditional Langmuir cells that appear in the absence of the farm, we employ a conditional sampling approach for the LES solutions to educe the coherent structure of both fields \citep[also see][]{McWiliams:1997jfm, Kukulka2010JPO, Roekel2012JGR, Shrestha2019EFM}. Based on the preconception of the form of cell structure, we identify the Langmuir cells by searching for the strong downwelling motion. The conditioning event $\mathscr{E}$ is defined as all $(x_s,y_s,t)$ instances that satisfy $w(x_s,y_s,z_{*},t) \leq -{\sigma_w}\big|_{max}$, where $\sigma_w$ is the RMS of transient vertical velocity and $z_{*}$ is the depth at which $\sigma_w$ attains its maximum value, denoted as ${\sigma_w}\big|_{max}$. The ordered pair $(x_s,y_s)$ represents a set of grid points in the horizontal space. For case LT, $\sigma_w =\langle \overline{w'w'} \rangle^{1/2}$ and $(x_s,y_s)$ enumerates the entire horizontal domain; while for case CLT, $\sigma_w =\langle \overline{w'w'} \rangle^{1/2}_y$ is a function of $x_s$, and $(x_s,y_s)$ only contains grid points at the center of the canopy spacing along the $x$-direction. Thus, the conditional average for any quantity, denoted as $\widehat{\phi}$, is obtained with,
\begin{equation} \label{condition}
\widehat{\phi}(x_s, y_s, x',y',z,t)=\left \langle \phi(x_s+x',y_s+y',z,t) \big| \mathscr{E}\right \rangle,\ 
\end{equation}
It should be noted that $(x, y)$ is the absolute coordinate in the horizontal plane based on the Cartesian system defined in figure \ref{fig:domain}, while $(x_s, y_s)$ denotes the reference point with $(x',y')$ being the distance from $(x_s, y_s)$ in the horizontal direction. Only when the flow is horizontally inhomogeneous should $(x_s, y_s)$ be equal to $(x, y)$. To reduce the sampling error, the sampled flow field for case CLT is then further smoothed by moving average with window size in the streamwise direction given by $x_s-h_\mathrm{b}/2 < x < x_s+h_\mathrm{b}/2$.

\begin{figure}
    \centering\includegraphics[width=1.0\linewidth]{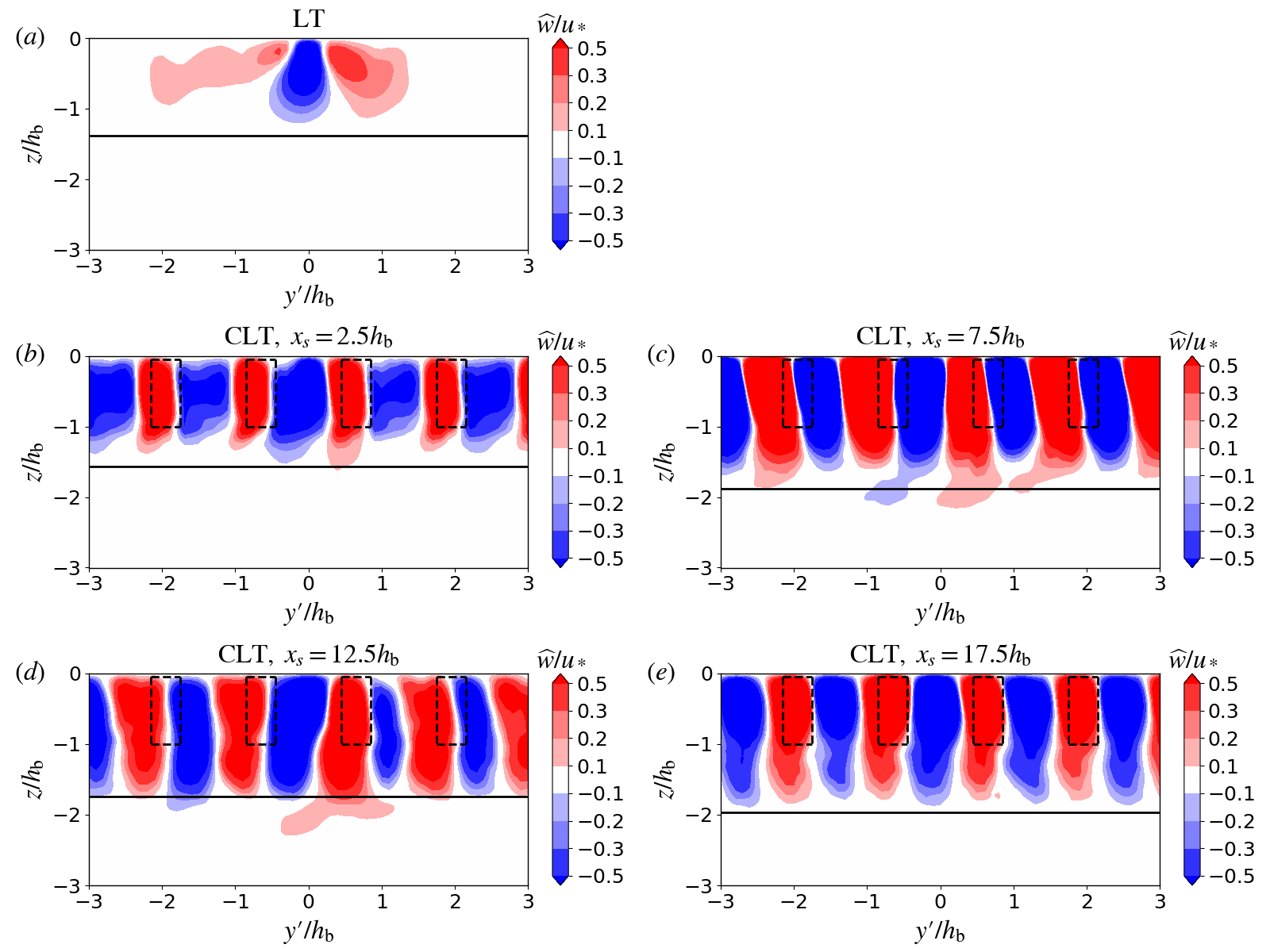} 
    \caption{Contour plots of the conditional-averaged transient vertical velocity $\widehat{w'}/u_{*}$ in $y-z$ planes for case LT (\textit{a}), and case CLT at different downstream locations (\textit{b}) $x_s=2.5h_\mathrm{b}$, (\textit{c}) $x_s=7.5h_\mathrm{b}$, (\textit{d}) $x_s=12.5h_\mathrm{b}$, and (e) $x_s=17.5h_\mathrm{b}$. The black solid line marks the mixed layer depth.}
\label{fig:cond_avg}
\end{figure}

Figure \ref{fig:cond_avg} shows the contour plots of $\widehat{w}/u_{*}$ in $y'-z$ planes for cases LT and CLT as noted in the caption. Note that the mean vertical velocity $\langle \overline{w} \rangle_y$ has been removed for the case CLT before conditional averaging operations to better compare the distinct attached Langmuir circulations against standard Langmuir circulations (e.g. $\langle \overline{w} \rangle_y$ is identically zero for LT but not for CLT). In both cases (LT and CLT), the Langmuir cells extend down to the bottom of the OML. The Langmuir cell pattern for LT (figure \ref{fig:cond_avg}\textit{a}) appears asymmetrical about the longitudinal plane because of the Ekman shear. The row spacing happens to be very close in width to the natural lateral size of the downwelling region in standard Langmuir circulations, and this may be related to the geometric characteristics of the attached Langmuir cells presented here. This canopy row spacing also plays a role in determining the separation between neighboring attached circulations as described above, and the effects of varying row spacing should be explored in the future. The downwelling velocity is greater than the upwelling velocity for both cases, but the upwelling motions increase by an order of magnitude in the presence of the canopy. This is partly caused by the fact that the obstruction of farm rows constrains the lateral extension of upwelling regions compared to standard Langmuir turbulence regime, producing stronger upwelling to conserve mass.

\section{Mechanism for attached Langmuir circulations} \label{sec:mechanism}

The standard Langmuir cells in a horizontally uniform OML (e.g. case LT) are generated through the CL2 instability, which is triggered by the wave-induced Stokes drift velocity acting upon a cross-stream perturbation in an otherwise horizontally uniform current \citep{Craik:1977JFM, Leibovich1983arfm, Suzuki2016JGR}. The instability arises from the torques produced by the variations of vortex force $\boldsymbol{u}_s \times \widetilde{\boldsymbol{\zeta}}$ that appears in \eqref{eq:momentum}, which leads to overturning cellular motions with downstream vorticity \citep{Leibovich1977jfm, Leibovich1983arfm}. This flow pattern drives the well-known Langmuir circulations that are transient in nature in the sense that they can survive for long periods of time but they also occasionally merge and disappear \citep{McWiliams:1997jfm}.

\begin{figure}
    \centering\includegraphics[width=0.7\linewidth]{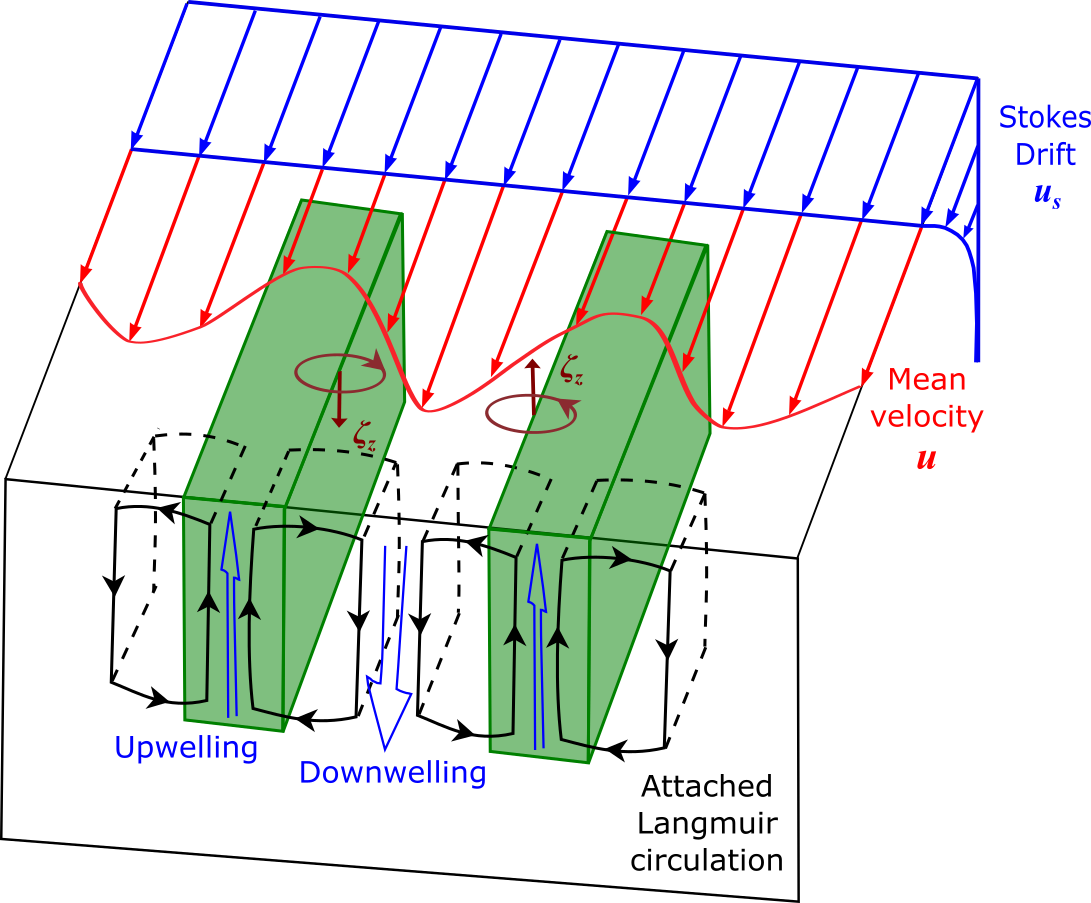}
    \caption{Sketch illustrating the mechanism for attached Langmuir circulations generated due to the presence of a farm in the upper ocean. The cross-varying current excited by the farm is rotated by the Stoke drift, producing the attached Langmuir circulations (black solid curves) that persist across the farm.}
\label{fig:mech}
\end{figure}

In the presence of a suspended farm with row structure (cases CLT and CLTL), the canopy drag acts within the fraction of volume occupied by the canopy elements, thus decelerating the fluid within the farm rows and accelerating the fluid in the spacing between rows due to continuity (figure \ref{fig:mech}). The cross-stream variation of the current produced by the farm generates a persistent vertical vorticity field $\zeta_z$ that interacts with the wave-induced Stokes drift in a way similar to the CL2 instability. Specifically, the vertical component of vorticity $\zeta_z$ associated with the cross-stream anomaly introduces a cross-stream vortex force $-u_s\zeta_z$ that carries fluid parcels towards the planes of local maximum $u$ where fluid sinks due to continuity \citep{Leibovich1983arfm, Thorpe2004ARFM}. Because the horizontal shear is persistent within the farm and the Stoke drift associated with the waves is horizontally uniform, such interaction gives rise to the formation of attached Langmuir circulations that are stationary and stable within the farm. This leads to downwelling regions in the high velocity regions between the canopy rows and upwelling regions within the rows of canopy elements. A schematic diagram illustrating the generation of such circulations is shown in figure \ref{fig:mech}. The black closed curves provide an illustration of the swirling streamlines in the plane perpendicular to the canopy axis. 

\begin{figure}
  \centering\includegraphics[width=1.0\linewidth]{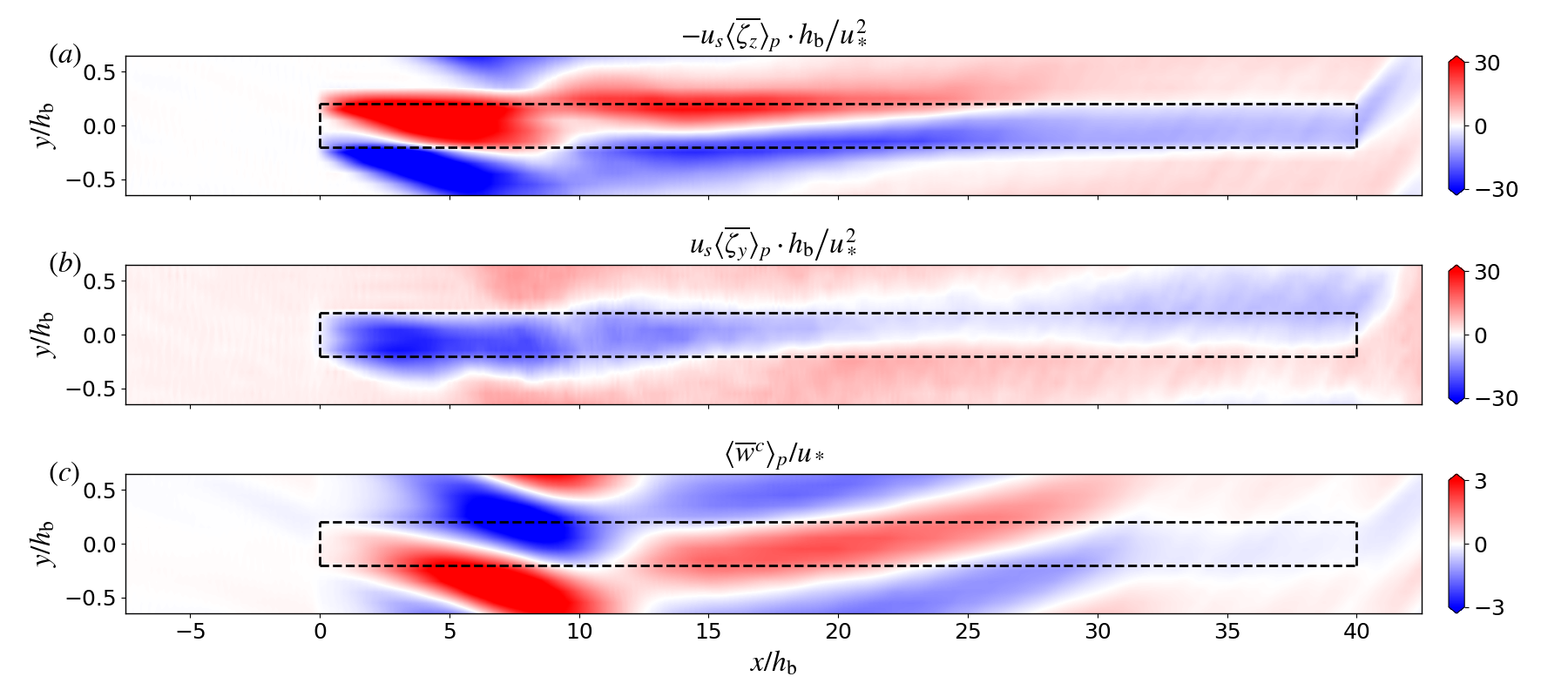}
  \caption{The time- and cross-phase-averaged vortex force: (\textit{a}) cross-stream component $-u_s\langle \overline{\zeta_z} \rangle_p \cdot h_\mathrm{b}/u_*^2$ and (\textit{b}) vertical component $u_s\langle \overline{\zeta_y} \rangle_p \cdot h_\mathrm{b}/u_*^2$; and (\textit{c}) secondary-flow part of vertical velocity $\langle \overline{w}^c \rangle_p/u_{*}$ for case CLTL at $z=-0.5h_\mathrm{b}$}.
\label{fig:vortex_force}
\end{figure}

Figures \ref{fig:vortex_force}\textit{a} and \textit{b} show the cross-stream and vertical components of vortex force, i.e. $-u_s\langle \overline{\zeta_z} \rangle_p$ and $u_s\langle \overline{\zeta_y} \rangle_p$ respectively, in the $x-y$ plane at $z=-0.5h_\mathrm{b}$ for the CLTL case. In terms of magnitude, the cross-stream component $-u_s\langle \overline{\zeta_z} \rangle_p$ dominates over the vertical component $u_s\langle \overline{\zeta_y} \rangle_p$ down to about $x/h_\mathrm{b} \le 30$, while they are both negligibly small towards the end of the longer farm. Consistent with the pattern of the coherent part of vertical velocity $\langle \overline{w}^c \rangle_p/u_{*}$ (figure \ref{fig:vortex_force}c), the vortex force alternates in sign periodically across the farm, forming pairs of equal magnitude, oppositely directed forces in the cross-stream direction. Very close to the leading edge ($0<x/h_\mathrm{b}<2$), as the flow just enters the farm, $-u_s\langle \overline{\zeta_z} \rangle_p$ is positive (pointing in the positive $y-$direction) and negative (pointing in the negative $y-$direction) near the left and right edges of the canopy rows, respectively. In consequence, the action of $-u_s\langle \overline{\zeta_z} \rangle_p$ drives upwelling motions within the farm rows and downwelling motions in the spacing (see figure \ref{fig:vortex_force}\textit{c}), as illustrated in figure \ref{fig:mech}. This pattern is clearly disrupted downstream from the leading edge, as discussed below.

\begin{figure}
  \centering\includegraphics[width=1.0\linewidth]{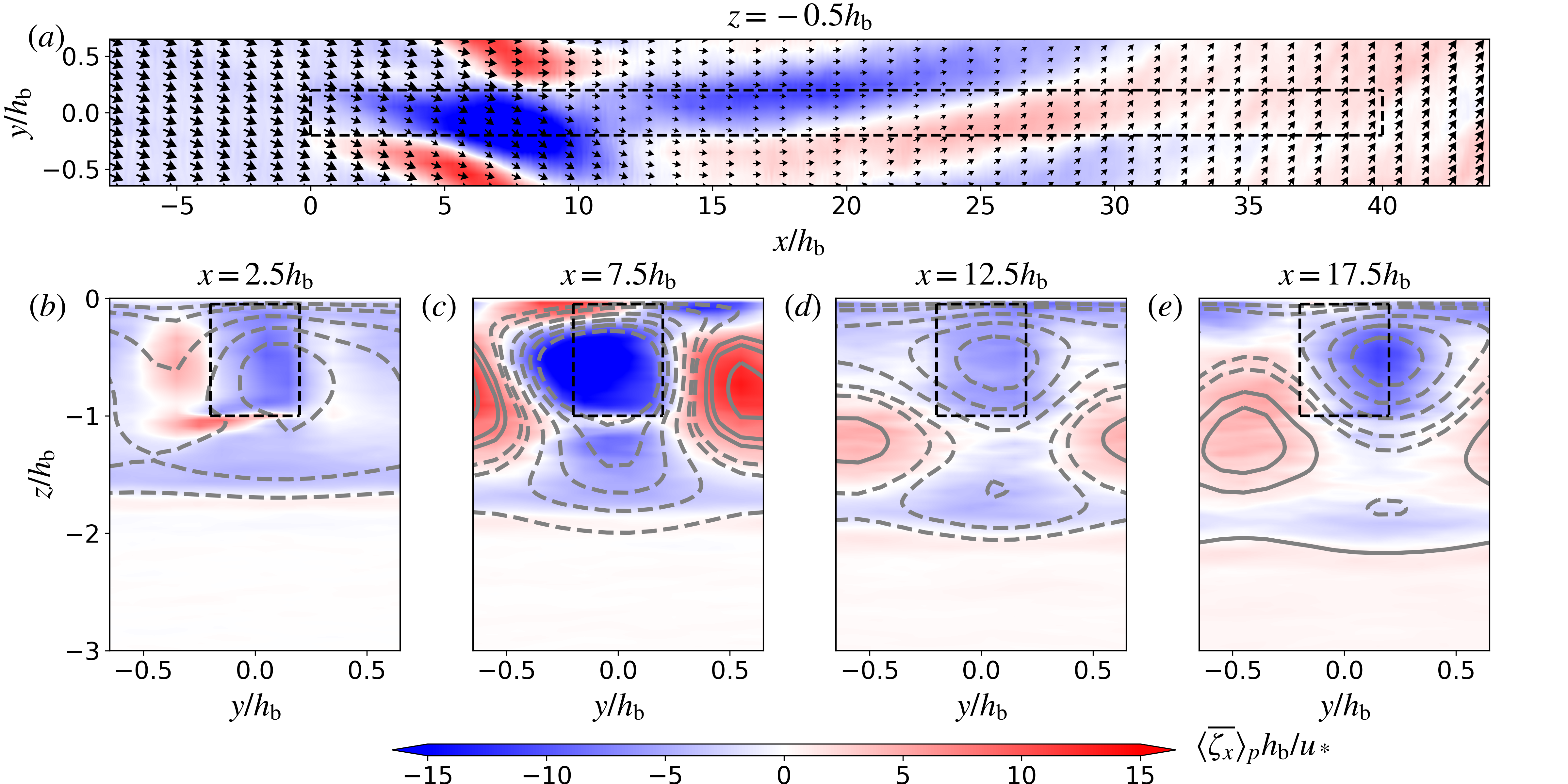}
  \caption{The time- and cross-phase-averaged downstream vorticity $\langle \overline{\zeta_x} \rangle_p h_\mathrm{b}/u_*$ with overlaid horizontal velocity vector (a scale factor of 1/5 is applied to $\langle \overline{u} \rangle_p$ for better visualization) for case CLTL at $z=-0.5h_\mathrm{b}$ (\textit{a}), and in the $y-z$ plane at four different downstream locations (\textit{b}) $x=2.5h_\mathrm{b}$, (\textit{c}) $x=7.5h_\mathrm{b}$, (\textit{d}) $x=12.5h_\mathrm{b}$, and (\textit{e}) $x=17.5h_\mathrm{b}$, overlying the two-dimensional streamfunction $\langle \overline{\psi} \rangle_p$ (grey lines) computed from $\zeta_x$.}
\label{fig:omega_x}
\end{figure}

To further characterize the flow structure associated with the attached Langmuir circulation, we look at the  streamwise vorticity $\zeta_x$. Figure \ref{fig:omega_x} plots the contours of $\langle \overline{\zeta_x} \rangle_p$ for case CLTL at several cross-sections as noted in the caption. As described above, it is the Stokes drift rotation of vertical vorticity $\zeta_z$ that produces downstream vorticity $\zeta_x$ of alternating signs in the cross-stream direction. Although the heterogeneous canopy in the absence of the Stokes drift (case CST) also generates turbulence-driven secondary flows (because of spatial variability of the turbulent stresses), it fails to yield any regular patterns in the streamwise vorticity as those shown in Figure \ref{fig:omega_x} (not shown). 

To better visualize the overturning circulations, we plot streamlines on $y-z$ cross-sections in figure \ref{fig:omega_x}\textit{b}-\textit{d} \citep{akselsen2019JFM, akselsen2020JFM}. We determine the streamlines as isolines of the non-divergent two-dimensional streamfunction $\psi$ computed from
\begin{equation} \label{eq:streamfunction}
    \frac{\partial^2 \psi}{\partial y^2} + \frac{\partial^2 \psi}{\partial z^2} = -\zeta_x, 
\end{equation}
The streamlines in figure \ref{fig:omega_x} portray pairs of counter-rotating vortices, with the axes aligned to the right of the wind for $0 < x/h_\mathrm{b} < 10$ and tilted to the left of the wind after $x/h_\mathrm{b} \approx 10$. Since the attached Langmuir cells are not strictly aligned with the $x-$direction, the use of $\langle \zeta_x \rangle_p$ only captures the largest downstream component of the three-dimensional vortices, and thus documents weaker overturning motions relative to the full form of coherent circulations. The variations of $\langle \overline{\zeta_x} \rangle_p$ resemble that of the secondary-flow part of vertical velocity in figure \ref{fig:w_se}\textit{a}, with the maximum magnitude appearing in the nearfield downstream from the leading edge ($x=2.5h_\mathrm{b} \sim 7.5h_\mathrm{b}$). Towards the end of farm, the negative downstream vortices vanish and only the weak positive vortices are left. This is mainly because the cross-stream vortex force therewith is not strong enough (figure \ref{fig:vortex_force}\textit{a}) to sustain a downstream counter-rotating vortex pair. 

In an idealized configuration in which the incoming mean flow is perfectly parallel to the farm rows, we would expect an organized flow structure similar to that shown in figure \ref{fig:mech}. However, as it is clearly seen in figures \ref{fig:w_se}, \ref{fig:vortex_force}, and \ref{fig:omega_x}, the patterns that emerge from the simulation are far more complex. The attached Langmuir cells meander in the cross-stream direction and their amplitude changes in a non-monotonic way as a function of distance from the leading edge of the farm. These departures from the idealized scenario are mostly caused by the cross-stream advection, as seen by the superposition of horizontal velocity vectors onto the streamwise vorticity in figure \ref{fig:omega_x}a. In particular, the shift in cross-stream velocity from negative to positive around $x/h_\mathrm{b}\approx15$ discussed in section \ref{sec:standing} produces a similar change in the effect of advection, causing the upwelling motions to be displaced to the right of the farm row in the region near the leading edge (i.e., up to $x/h_\mathrm{b}\approx10$) and to the left of the row for $x/h_\mathrm{b}>18$.

The same pattern observed in the upwelling/downwelling regions is clearly seen in the streamwise vorticity, as the two quantities are tied together by the overturning structure of the flow. However, the advection of the vertical and cross-stream components of vorticity is less effective, as clearly seen in the patterns of the vortex force (which reflect the patterns of $\langle \overline{\zeta_z} \rangle$ and $\langle \overline{\zeta_y}\rangle$). This is mostly because the canopy drag continues to generate lateral shear at the canopy edges, strongly influencing the position of $\langle \overline{\zeta_z} \rangle$ and $\langle \overline{\zeta_y}\rangle$. As a consequence, in the region between $10<x/h_\mathrm{b}<15$, the upwelling/downwelling branches of the attached Langmuir cells no longer coincide with the divergence/convergence of the cross-stream vortex force (compare figures \ref{fig:vortex_force}\textit{a} and \ref{fig:vortex_force}\textit{c}), leading to the weakening of the attached Langmuir cells around $x/h_\mathrm{b}=12$ followed by a restrengthening at the more favorable position with the upwelling within the canopy row. This process appears mostly as an abrupt left shift of the flow structure at $x/h_\mathrm{b}\approx12$. 
Towards the end of the farm, $-u_s\langle \overline{\zeta_z} \rangle_p$ is significantly reduced, and is no longer capable of driving clear attached Langmuir circulations (see figures \ref{fig:vortex_force}\textit{c} and \ref{fig:omega_x}), which is also consistent with the decay of the vertical variance for the secondary-flow component of the flow seen in figure \ref{fig:wrms_line_comp}.

\section{Mixed layer energetics} \label{sec:energetics}

In this section, we examine the budget of the kinetic energy in the mixed layer, which will reveal the energy source for the secondary flow in our LES solutions. Following the decomposition strategy described in section \ref{sec:decomposition}, the total kinetic energy ($K=\left<\overline{u_iu_i}\right>_y/2$) is composed of contributions due to the mean flow, secondary flow, and transient eddies as,
\begin{equation}  \label{eq:totalKE}
    K=\frac{1}{2}\left<\overline{u_iu_i}\right>_y = 
    \underbrace{ \vphantom{ \left<\overline{u_i'u_i'}\right>_y }
    \frac{1}{2}\left<\overline{u_i}\right>_y\left<\overline{u_i}\right>_y}_{K_\mathrm{M}} +
    \underbrace{ \vphantom{ \left<\overline{u_i'u_i'}\right>_y }
    \frac{1}{2}\left<\overline{u_i}^c \overline{u_i}^c\right>_y}_{K_\mathrm{SE}} +
    \underbrace{\frac{1}{2}\left<\overline{u_i'u_i'}\right>_y}_{K_\mathrm{TE}}
\end{equation}
Here, $K_\mathrm{M}$ represents the mean kinetic energy, $K_\mathrm{SE}$ is the kinetic energy of the secondary mean flow (which includes lateral variations in the flow produced by the spatial structure of the farm and the attached Langmuir circulations), and $K_\mathrm{TE}$ is the turbulent kinetic energy. By manipulating the governing equations \eqref{eq:continuity} and \eqref{eq:momentum}, the transport equations for $K_\mathrm{M}$, $K_\mathrm{SE}$ and $K_\mathrm{TE}$ can be obtained as follows, 
\begin{subequations} \label{eq:budget}
\begin{align}
  \frac{\mathrm{D}{K_\mathrm{M}}}{\mathrm{D}{t}} & =
    -C_\mathrm{M-SE}-C_\mathrm{M-TE} + S_\mathrm{M} + B_\mathrm{M} +
    \varepsilon_\mathrm{M} + D_\mathrm{M} + T_\mathrm{M} + R_\mathrm{M} \label{eq:budget_a}, \\[3pt]
  \frac{\mathrm{D}{K_\mathrm{SE}}}{\mathrm{D}{t}} & = 
    C_\mathrm{M-SE} - C_\mathrm{SE-TE} + S_\mathrm{SE} + B_\mathrm{SE} +
    \varepsilon_\mathrm{SE} + D_\mathrm{SE} + T_\mathrm{SE} \label{eq:budget_b}, \\[3pt]
  \frac{\mathrm{D}{K_\mathrm{TE}}}{\mathrm{D}{t}} & = 
    C_\mathrm{M-TE} + C_\mathrm{SE-TE} + S_\mathrm{TE} + B_\mathrm{TE} + 
    \varepsilon_\mathrm{TE} + D_\mathrm{TE} +T_\mathrm{SE} \label{eq:budget_c}. 
\end{align}
\end{subequations}
in which the material derivative $\mathrm{D}/\mathrm{D}{t}=\partial/\partial{t} + \left<\overline{u_j}\right>_y \partial/\partial{x_j} + u_s\partial/\partial{x}$. Note that the prescribed wave and current conditions, namely $\boldsymbol{u}_{g}=(u_g, 0, 0)$ and $\boldsymbol{u}_{s}=(u_s(z), 0, 0)$, have been invoked in deriving these equations. The first two terms on the RHS of \eqref{eq:budget} represent the magnitude of energy conversion between $K_\mathrm{M}$, $K_\mathrm{SE}$, and $K_\mathrm{TE}$ as implied in the subscripts, and are given by,
\begin{equation}  \label{eq:conversion}
\left. \begin{array}{l}
\displaystyle 
    C_\mathrm{M-SE} = -\left<\overline{u_i}^c \overline{u_j}^c\right>_y \frac{\partial{\left<\overline{u_i}\right>_y}}{\partial{x_j}}, \\[16pt]
\displaystyle
    C_\mathrm{M-TE} = -\left<\overline{u_i'u_j'}\right>_y \frac{\partial{\left<\overline{u_i}\right>_y}}{\partial{x_j}}, \\[16pt]
\displaystyle
    C_\mathrm{SE-TE} = -\left<\overline{u_i'u_j'}\frac{\partial{\overline{u_i}^c}}{\partial{x_j}}\right>_y.
\end{array} \right\}
\end{equation}
Note that the Einstein summations convention is used. As an example, $C_\mathrm{M-SE}>0$ represents the rate of production of $K_\mathrm{SE}$ at the expense of $K_\mathrm{M}$, as this term appears as a source in the equation for $K_\mathrm{SE}$ \eqref{eq:budget_b} and a sink in the equation for $K_\mathrm{M}$ \eqref{eq:budget_a}. Thus, it represents the energy transfer rate from the mean flow to the secondary flow.

The third terms on the RHS of equations \eqref{eq:budget} are the Stokes production terms that reflect the energy conversion between the waves and the decomposed field,
\begin{equation}  \label{eq:Stokes_prod}
    S_\mathrm{M}=-\left<\overline{u}\right>_y\left<\overline{w}\right>_y\partial{u_s}/\partial{z}, \quad
    S_\mathrm{SE} = -\left<\overline{u}^c \overline{w}^c\right>_y \partial{u_s}/\partial{z}, \quad
    S_\mathrm{TE} = -\left<\overline{u'w'}\right>_y \partial{u_s}/\partial{z}
\end{equation}
Interestingly, the Stokes production, which only makes contribution to the turbulent kinetic energy in a horizontally homogeneous OML, now also appears in the budget equation of mean kinetic energy in our LES experiments because of a non-zero and spatially evolving mean vertical velocity $\left<\overline{w}\right>_y$ field. The fourth term is the buoyancy production term,
\begin{equation}  \label{eq:buoy_prod}
    B_\mathrm{M} = \alpha g \left<\overline{w}\right>_y \left( \left<\overline{\theta}\right>_y - \theta_0 \right), \quad
    B_\mathrm{SE} = \alpha g \left<\overline{w}^c \overline{\theta}^c\right>_y, \quad
    B_\mathrm{TE} = \alpha g \left<\overline{w'\theta'}\right>_y
\end{equation}
Here, $B_\mathrm{M}$ represents an exchange of mean kinetic energy $K_\mathrm{M}$ with the potential energy. The fifth term in \eqref{eq:budget} is the SGS dissipation term,
\begin{equation}  \label{eq:sgs_dissip}
    \varepsilon_\mathrm{M} = -\left<\overline{\tau_{ij}}\right>_y\partial{\left<\overline{u_i}\right>_y} / \partial{x_j}, \quad
    \varepsilon_\mathrm{SE} = -\left<\overline{\tau_{ij}}^c\partial{\overline{u_i}^c}/\partial{x_j}\right>_y, \quad
    \varepsilon_\mathrm{TE} = -\left<\overline{\tau_{ij}'\partial{u_i'}/\partial{x_j}}\right>_y
\end{equation}
In light of the energy cascade phenomenology \citep{Pope2000}, we expect most of the energy dissipation occurs at the small-scale transient eddies, while the energy loss of the large-scale mean flow and secondary flow to direct SGS dissipation effects is negligible, i.e. $\varepsilon_\mathrm{M},\ \varepsilon_\mathrm{SE} \ll \varepsilon_\mathrm{TE}$. Thus, we will assume $\varepsilon \approx \varepsilon_\mathrm{TE}$ in interpreting the LES solutions, and do not partition the total dissipation $\varepsilon$ into 3 components as in \eqref{eq:sgs_dissip}. 
The sixth term in \eqref{eq:budget} is the canopy destruction term,
\begin{equation}  \label{eq:drag_prod}
    D_\mathrm{M} = - \left<\overline{u_i}\right>_y\left<\overline{F_{D,i}}\right>_y, \quad
    D_\mathrm{SE} = - \left<\overline{u_i}^c \overline{F_{D,i}}^c\right>_y, \quad
    D_\mathrm{TE} = - \left<\overline{u_i'F_{D,i}'}\right>_y   
\end{equation}
which represents the energy gain/loss of each component of the flow field (i.e. mean flow, secondary flow, and transient eddies) due to the action of canopy drag. The terms in flux form are grouped together as a transport term in \eqref{eq:budget},
\begin{subequations}
\begin{align}
  \begin{split}
    T_\mathrm{M} &= \frac{\partial{}}{\partial{x_j}} \left[  \left<\overline{u_i}\right>_y\left<\overline{\tau_{ij}}\right>_y+\left<\overline{u_j}\right>_y\left<\overline{u}\right>_y u_s -\left<\overline{u_j}\right>_y\left<\overline{\Pi}\right>_y \right.\\
    &\quad\quad\quad\quad{} \left. -\left<\overline{u_i}\right>_y\left<\overline{u_i}^c \overline{u_j}^c\right>_y -\left<\overline{u_i}\right>_y\left<\overline{u_i'u_j'}\right>_y \right], \\
  \end{split} \\
  T_\mathrm{SE} &= \frac{\partial{}}{\partial{x_j}} \left< \overline{u_i}^c \overline{\tau_{ij}}^c - \frac{1}{2} \overline{u_i}^c \overline{u_i}^c \overline{u_j}^c -\overline{u_j}^c \overline{\Pi}^c-\overline{u_i'u_j'}\overline{u_i}^c+\overline{u_j}^c \overline{u}^c u_s \right>_y, \\
   T_\mathrm{TE} &= \frac{\partial{}}{\partial{x_j}} \left< \overline{u_i'\tau_{ij}'} - \overline{u_i'u_i'u_j'}/2-\overline{u_j'\Pi'} + \overline{u_j'u'}u_s-\frac{1}{2} \overline{u_j}^c \overline{u_i'u_i'} \right>_y.
\end{align} 
\end{subequations}
which represents the transport of kinetic energy ($K_\mathrm{M}$, $K_\mathrm{SE}$, or $K_\mathrm{TE}$) through resolved momentum stresses, SGS stresses, and pressure. The last term on the RHS of \eqref{eq:budget_a} represents the effect of Coriolis force associated with the Stokes drift and geostrophic current, 
\begin{equation}  \label{eq:coriolis-stokes}
R_\mathrm{M} = f\left<\overline{v}\right>_y(u_g-u_s) 
\end{equation}
which transfers energy from surface waves and external larger-scale field to the mean flow \citep{Suzuki2016JGR}.

\begin{figure}
    \centering\includegraphics[width=1.0\linewidth]{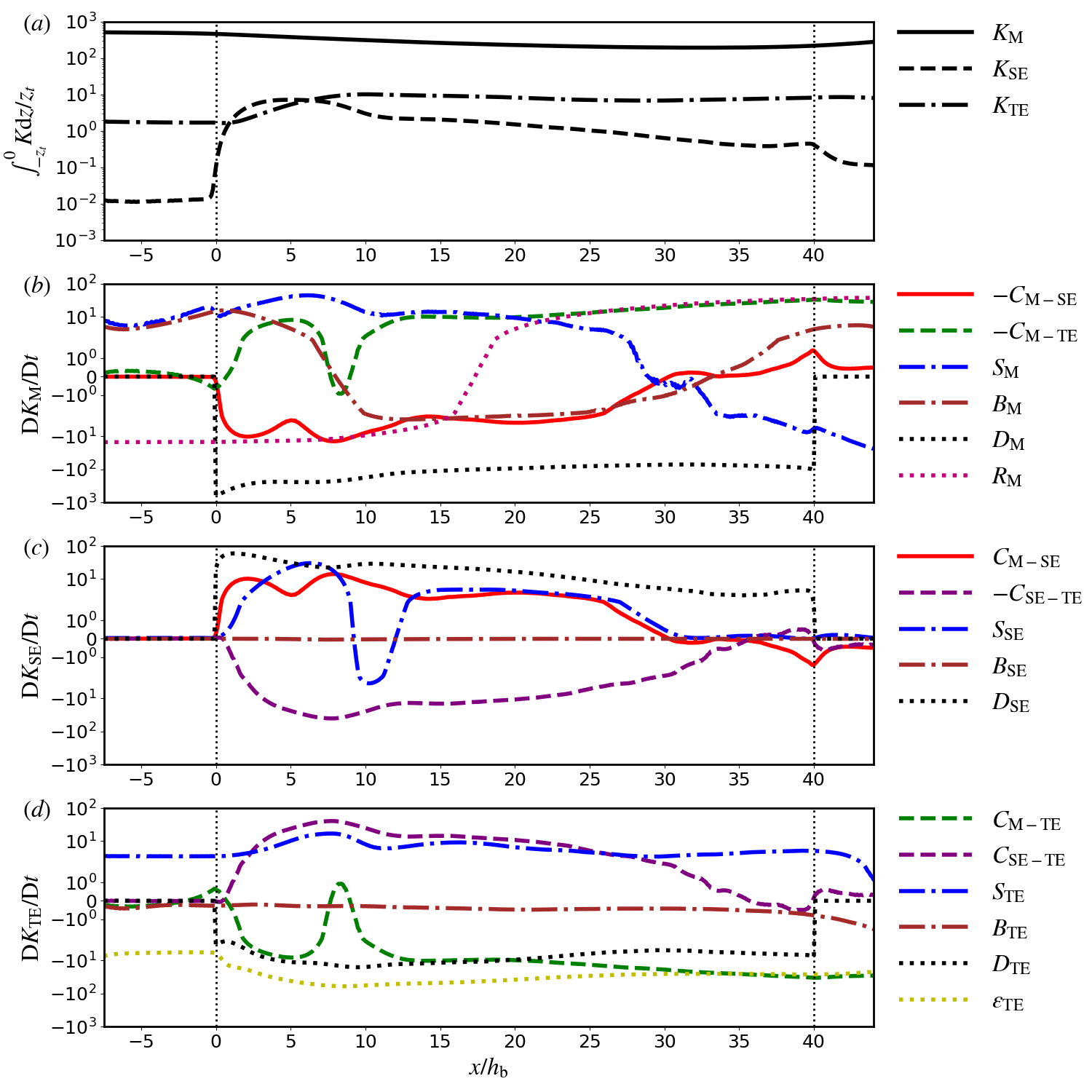}
    \caption{Budget terms of the depth-averaged kinetic energy in the upper surface layer for case CLTL: (a) downstream variation of the triply-decomposed kinetic energy; and partition of conversion, Stokes production, buoyancy production and canopy destruction for (b) $K_\mathrm{m}$, (c) $K_\mathrm{SE}$, and (d) $K_\mathrm{TE}$. The terms are normalized by $h_\mathrm{b}/u_*^{3}$.}
\label{fig:ke_budget}
\end{figure}

\begin{figure}
    \centering\includegraphics[width=0.8\linewidth]{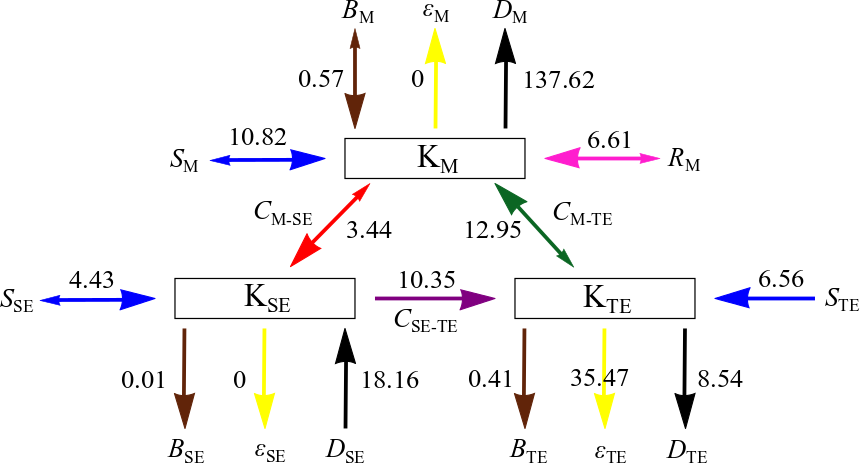}
    \caption{Schematic diagram of the depth-averaged energy budget for the mean flow, secondary flow, and transient eddies. The arrow-lines represent the transfer of energy integrated over the entire farm length, with the direction of net energy flow indicated by heavy arrowheads. The number alongside each arrow-line is the farm-averaged value of the corresponding term, normalized by $h_\mathrm{b}/u_*^{3}$. Note that the transport terms are not included here, thus the energy budget for each component is not closed.}
\label{fig:ke_relation}
\end{figure}

Figure \ref{fig:ke_budget}\textit{a} shows the downstream variation of the depth-averaged kinetic energy for the triply-decomposed field. Within the canopy region ($0 < x/h_\mathrm{b}<40$), $K_\mathrm{M}$ decreases because the farm drains the mean kinetic energy by decelerating the time-mean flow. As the OML flow impinges upon the farm, both $K_\mathrm{SE}$ and $K_\mathrm{TE}$ increase in the near-field downstream of the leading edge. While $K_\mathrm{TE}$ maintains at a high level after that, $K_\mathrm{SE}$ gradually decreases towards the end of the farm. This suggests that, in the presence of a suspended farm, the flow within the canopy region is in a highly turbulent state but the organized secondary circulations become less intense as the fluid moves further downstream. The downstream variations of the various production and destruction terms in the kinetic energy budget equation \eqref{eq:budget} for the mean flow, secondary flow, and transient eddies are depicted in figure \ref{fig:ke_budget}\textit{b}-\textit{d}, respectively, using $u_{*}$ and $h_\mathrm{b}$ as the scaling parameters (transport terms are not shown). To facilitate interpretation, the curves are color coded according to the diagram depicting energy exchanges shown in figure \ref{fig:ke_relation}, which provides a summary of the energy budget for the three components of the flow integrated over the entire farm. 

The Stokes production $S_\mathrm{M}$ is the main source for $K_\mathrm{M}$ (figure \ref{fig:ke_budget}\textit{b}), except it is negative after $x/h_\mathrm{b} \approx 32$, mainly because of the upward deflection near the trailing edge, i.e. $\left<\overline{w}\right>_y > 0$ that makes $S_\mathrm{M}=-\left<\overline{u}\right>_y\left<\overline{w}\right>_y\partial{u_s}/\partial{z}<0$. Contrary to expectations, the energy conversion term $-C_\mathrm{M-TE}$ is mostly positive along the farm (green dashed line in figure \ref{fig:ke_budget}\textit{b} and \textit{d}), indicating that the transient eddies lose kinetic energy to the mean flow. The canopy destruction term $D_\mathrm{M}$ is the primary sink term for $K_\mathrm{M}$ as the hydrodynamic drag imparted by the farm consistently removes the momentum from the flow (e.g. ${\partial{\left<\overline{u}\right>_y}}/{\partial{x}}<0$). The energy conversion term $C_\mathrm{M-SE}$ (red solid line in figure \ref{fig:ke_budget}\textit{b} and \textit{c}) constitutes the secondary energy sink for $K_\mathrm{M}$, i.e. energy is transferred from the mean flow to the secondary flow. This is mainly because the leading order term of $C_\mathrm{M-SE}$ in \eqref{eq:conversion} is $-\left<\overline{u}^c \overline{u}^c\right>_y {\partial{\left<\overline{u}\right>_y}}/{\partial{x}}>0$. Since the geostrophic current and Stokes drift velocity are prescribed, the sign of Coriolis-related term $R_\mathrm{M}$ in \eqref{eq:coriolis-stokes} is directly determined by the cross-stream velocity $\left<\overline{v}\right>_y$, which goes to the right of the wind (i.e. $\left<\overline{v}\right>_y<0$) as in standard Langmuir turbulence before $x/h_\mathrm{b} \approx 18$ and then turn to the left of the wind (i.e. $\left<\overline{v}\right>_y>0$) after that (not shown). The flow veering is largely caused by the modification of the suspended farm on the vertical momentum transfer, given that $f\left<\overline{v}\right>_y \sim \partial{\left<\overline{u'w'}\right>_y}/\partial{z}$ as yielded from a reduced form of \eqref{eq:momentum}.

In terms of the secondary flow, the canopy-related term $D_\mathrm{SE}$ is a major source term for $K_\mathrm{SE}$ (black dotted line in figure \ref{fig:ke_budget}\textit{c}), mainly because it is the spatial arrangement of the farm that leads to persistent variations in the streamwise flow across the farm. Apart from the energy conversion from the mean flow $C_\mathrm{M-SE}$, another important source term for the secondary mean flow is the Stokes production $S_\mathrm{SE}$, which is the main source of energy to the attached Langmuir circulations. This is true everywhere except for the region $9 < x/h_\mathrm{b} < 12$ where $S_\mathrm{SE}$ is negative. In this local range, $S_\mathrm{SE}$ serves as an sink of $K_\mathrm{SE}$ and the energy transferred from the mean flow $C_\mathrm{M-SE}$ is also decreasing (red solid line), which to some extent explains the local attenuation of attached Langmuir circulations at $x/h_\mathrm{b}=12.5$ (figure \ref{fig:w_se}d). For $x/h_\mathrm{b} > 32$, $S_\mathrm{SE}$ is approximately zero because the coherent vertical velocity $\overline{w}^c$ almost vanishes (figure \ref{fig:wrms_line_comp}) and hence the momentum stress due to the secondary flow $\left<\overline{u}^c \overline{w}^c\right>_y$ in \eqref{eq:Stokes_prod} is negligibly small. These three source terms ($D_\mathrm{SE}$, $S_\mathrm{SE}$, and $C_\mathrm{M-SE}$) are responsible for the maintenance of secondary flow (including the attached Langmuir circulations) in the adjustment region downstream of the leading edge, whereas the exchange with the transient eddies $C_\mathrm{SE-TE}$ constantly extracts energy from the secondary flow to support the turbulence level (purple dashed line in figure \ref{fig:ke_budget}c).

As shown in figure \ref{fig:ke_budget}\textit{d}, the transient eddies feed on wave energy transferred by the Stokes drift shear (blue dash-dotted line) and energy conversion from the secondary flow. The transient eddies lose energy mostly via three processes: (i) energy transfer to the mean flow; (ii) energy removal due to the canopy drag; and (iii) energy dissipation at the small scales (represented by the SGS dissipation). As Langmuir turbulence in the presence of canopy features strong shear layers and wave forcing, and we assume no incoming or outgoing buoyancy flux at the surface, the buoyant production terms for the secondary flow and transient eddies ($B_\mathrm{SE}$ and $B_\mathrm{TE}$) are negligibly small in comparison.

\section{Conclusions} \label{sec:conclusion}

In this study, a fine-scale LES model is used to explore how Langmuir turbulence in deep ocean evolves as it flows over and through a row-structured macroalgae farm. The ocean flow is driven by a constant wind stress and a geostrophic current, under the influences of surface gravity waves, planetary rotation, and stable interior stratification. The effects of Langmuir turbulence are accounted for by adding the CL vortex force to the momentum equation without explicitly resolving the surface waves. For the case studied here, the drag force at the bottom of the farm becomes dominant over the wind forcing at the surface with increasing distance downstream from the leading edge. As a result, the mean horizontal flow switches from the canonical surface forced Ekman layer to a regime that resembles a bottom Ekman layer. This transition is evident in the change of direction of the mean current perpendicular to the wind.

Following a triple-decomposition technique, the turbulent transport is divided into contributions from the mean flow, secondary flow, and transient fluctuations. We find out that the row structure of the farm causes the cross-stream variation of the current that ultimately leads to the formation of coherent circulations via a mechanism similar to CL2 instability theory. Specifically, the vertical vortex lines associated with this cross-varying current are tilted by the Stokes drift, driving the formation of downstream vortices that are stationary in time, phase locked in space, and periodically alternating in sign across the lateral direction. Thus, we also refer to these coherent structures as attached Langmuir circulations. 

The attached Langmuir circulations are unique to the upper OML in the presence of aquacultural farms (or other distributed roughness elements) since the cross-stream variation of the current is excited by the canopy. They are roughly oriented along the rows of canopy elements, which are aligned with the wind direction within the present numerical framework. The vertical extent of attached Langmuir circulations can occupy the entire OML, with the lateral scale of the associated downwelling regions comparable to the row spacing in the farm. The potential impact of varying farm geometry and ocean conditions on these circulations is out of the scope here, but should be explored in the future.

Because the associated upwelling motions are concentrated in regions occupied by macroalgae elements, these attached Langmuir circulations are conducive to vertical mixing and could increase nutrient availability within macroalgae farm environments. The strength of transient eddies, characterized by cross-stream and vertical turbulence intensity (i.e. $\langle \overline{v'v'} \rangle^{1/2}_y$ and $\langle \overline{w'w'} \rangle^{1/2}_y$), is much larger under the effect of Stokes drift associated with the surface waves (case CLT) compared to the pure shear-driven scenario (case CST), which is also consistent with previous studies in the absence of the canopy \citep{McWiliams:1997jfm, D'Asaro2001JPO}. For both cases, the suspended farm prompts a shear layer development near the canopy bottom and deepens the mixed layer as the flow moves downstream. For the simulation with Stokes drift (case CLT), in the nearfield downstream of the leading edge, the canopy drag dampens the turbulence, leading to the reduction of $\langle \overline{w'w'} \rangle^{1/2}_p/u_{*}$ for $0<x/h_\mathrm{b}<4$. Further downstream, the attached Langmuir circulations promote strong enhancement of turbulence. This enhancement slowly fades as the flow adjusts to the canopy and the strength in the secondary flow decays (figure \ref{fig:wrms_line_comp}). The presence of the canopy leads to the formation of the attached Langmuir circulations and to local enhancement of the turbulence. Both flow modifications are expected to enhance vertical mixing within the OML and possibly help the entrainment of nutrients from the pycnocline.

Analysis of kinetic energy budget shows that, as the flow moves downstream of the canopy leading edge, the canopy drag acts as an energy sink for the mean flow and transient fluctuations, while serving as a major source for the kinetic energy of the secondary mean flow. If the canopy is long enough, the secondary flow pattern vanishes when the oceanic turbulence is fully adjusted to the macroalgal farm. Therefore, this flow feature arises from the adjustment of the upper mixed layer to the aquafarm.

The conclusions drawn here are valid for conditions in which the effect of Stokes drift dominates over that of wind stress and external pressure gradient forcing (i.e. the solutions are posed in the Langmuir turbulence regime). Despite the simplification made here (e.g. plant reconfiguration, monochromatic waves, etc.), we are optimistic that the findings presented above are relevant to realistic practice, and could serve as guidance for the design of large scale macroalgae systems. Still, the attached Langmuir circulations from our LES solutions and their potential implication on nutrient uptake by aquaculture farms await field observations to confirm their veracity.

From a fluid dynamics perspective, the physical flow presented here encompasses a variety of processes (stratification, Coriolis acceleration, wave-driven transport, and a canopy, etc). One of our main goals is to make it clear that these flow features are important in practice, in conditions under which macroalgal farms are deployed. As it turns out, most of the complexity involved in our setup is essential for the attached Langmuir eddies to develop (waves, mean current, non-uniform canopy, and downstream flow development). There are some possible simplifications that would allow one to reduce the parameter space and simplify the problem, bringing it to a more manageable fundamental configuration (e.g., removing the effects of planetary rotation and stratification). The results in this paper warrant further investigation of a more fundamental nature in simplified conditions, which could help reconcile a bit the complexity of the flow features we discovered with a more traditional fluid dynamical investigation of the parameter space.
\\

\section*{Acknowledgements}
This work is supported by the ARPA-E MARINER Program (DE-AR0000920). We thank the three anonymous reviewers for their constructive comments which led to improvements of the manuscript.

\section*{Declaration of Interests}
The authors report no conflict of interest.

\appendix
\renewcommand{\thesection}{\Alph{section}}

\section{Motion of buoyant, flexible macroalgae in upper OML}\label{appA}
The stipe reconfiguration in response to the flowing water depends on the ocean parameters (wave amplitude, wave period, and current) and the mechanical properties of macroalgae (stipe length, Young's modulus, density, and buoyancy). We decompose the upper OML flow into two parts, i.e. the steady flow (geostrophic current) and oscillatory flow (wave orbital velocity), and analyze the motion of buoyant, flexible macroalgae with respect to flow components separately. For each plant, the stipe bundles are simplified to have a circular cross-section, with length $l_s=20$ m, radius $r_s=0.1$ m (corresponding second moment of area $I=\upi r_s^4/4$), Young's modulus $E=1 \times 10^7$ Pa, and density $\rho_\mathrm{s}=595\ \mathrm{kg\ m^{-3}}$ \citep[properties taken from][]{Utter:1996,Henderson2019ce}.

In a unidirectional steady current (e.g. $u_g = 0.2\ \mathrm{m\ s^{-1}}$), the key parameters determining the form of macroalgae elements in sustained flow conditions are the dimensionless Cauchy number $\textit{Ca}$ (fluid drag/elastic force) and buoyancy number $\textit{B}$ (buoyancy force/elastic force) defined as \citep{Luhar2011lo}, 
\begin{equation} \label{eq:Ca}
    Ca=\frac{1}{2}\frac{\rho C_D r_s l_s u_g^2 }{EI/l_s^2}, 
\end{equation}
\begin{equation} \label{eq:B}
    B=\frac{\left( \rho-\rho_s \right) g \upi r_s^2 l_s}{EI/l_s^2}, 
\end{equation}
in which $\rho=1010\ \mathrm{kg\ m^{-3}}$ is the density of water. The ratio $Ca/B$ measures the relative importance of fluid drag and buoyancy force. Note that the flexibility of blades is neglected, because the blades can fold and rotate in the water, while the stipe bundles constitute the essential part governing the bending of macroalage elements. As such, only the fluid drag on the stipe bundles (denominator in \eqref{eq:Ca}) is considered instead of \eqref{eq:drag}. From the values of parameters given above, the resulting $Ca/B = 2.3 \times 10^{-3} \ll 1\ (Ca = 304.5,\ B = 1.3 \times 10^5)$, and the bending angle of the stipe bundles $\xi = 0.13\degree$ \citep[estimated by equation (12) in][]{Luhar2011lo}, suggesting the buoyancy force dominates over the fluid drag and the stipe bundles deform very little relative to its vertical position.

For wave-induced oscillatory flows, such as a sinusoidal wave with surface elevation $\eta = a_w \cos{\left( kx-\sigma t_w \right )}$, \citet{Henderson2019ce} introduced a new dimensionless buoyancy number $\beta$ and stiffness number $S$,
\begin{equation} \label{eq:beta}
    \beta=\frac{\left( \rho-\rho_s \right) g r_s t_w}{\rho C_D l_s u_w}, 
\end{equation}

\begin{equation} \label{eq:S}
    S=\frac{E I t_w}{\rho C_D r_s l_s^4 u_w}, 
\end{equation}
Here, $\sigma=2\upi/a_w$ is the angular frequency, $t_w=2\upi/\sqrt{gk}$ and $u_w=\sigma a_w$ are the wave period and orbital velocity scale, respectively. Based on the monochromatic wave parameters reported in section \ref{sec:simulation} ($t_w=6.2$ s, $u_w=0.81\ \mathrm{m\ s^{-1}}$), the resulting $\beta=1.06$ and $S=3.2 \times 10^{-5}$. The relative magnitude of buoyancy and elasticity scales with $\gamma=\beta/S^{1/2}=184 \gg 1$, which suggests that the elasticity plays a negligible role here. As $\beta$ is of order unity, the stipe displacement and the wave-induced water motion are comparable, i.e. the stipe bend with the waves. Note that our estimates of $S$ and $\beta$ are different from those in \citet{Henderson2019ce} because different values of wave (e.g. period and amplitude) and canopy parameters (e.g. length and drag coefficient) are used here.

\section{Deep-water wave attenuation by suspended canopies}\label{appB}
Surface waves propagating through marine plants lose energy due to the drag exerted by the canopy, leading to attenuation in wave heights \citep{Dalrymple1984}. Canopy configuration (subemerged, emergent, suspended) and the associated spatial distribution patterns exert a major impact on wave attenuation \citep{Chen2019AWR}. The following mathematical derivation is based on the work of \citet{Dalrymple1984} for damping by rigid cylinders in coastal regions, and considers suspended macroalgal farms in deep water (described previously in the main text).

Assuming that energy dissipation is dominated by the canopy drag force, the conservation of wave energy equation is, 
\begin{equation} \label{eq:wave_energy}
    \frac{\partial \left( E_w c_g \right)}{\partial x}= -\alpha_D \varepsilon_D, 
\end{equation}
in which $E_w=\frac{1}{2}\rho g a_w^2$ is the energy density per unit area of sea surface waves, $a_w$ is the wave amplitude, and $c_g = \frac{1}{2}\sqrt{g/k}$ is the wave group velocity. The prefactor $\alpha_D$ accounts for the reduction in dissipation arising from the motion of buoyant, flexible macroalgae, and it is a function of $\beta$ and $S$ defined in appendix \ref{appA} expressed as \citep{Henderson2019ce}, 
\begin{equation} \label{eq:wave_prefactor}
    \alpha_D= \left[ \frac{C_S S + C_\beta \beta^2}{1 + C_S S + C_\beta \beta^2}  \right]^{1/4}, 
\end{equation}
in which $C_S=1/4$ and $C_\beta=1/16$. For the highly flexible macroalgae ($\beta=1.06$ and $S=3.2 \times 10^{-5}$ in appendix \ref{appA}), the value of $\alpha_D$ is 0.51. $\varepsilon_D$ is the mean depth-integrated wave dissipation due to canopy drag force, 
\begin{equation} \label{eq:wave_dissipation}
    \varepsilon_D = \overline{\int_{-h_b}^{0} D_x u_x\ \mathrm{d} z},
\end{equation}
in which the overline denotes averaging over a complete wave period, $u_x = \sigma a e^{kz} \cos{\left( kx-\sigma t \right)}$ is the horizontal velocity due to wave orbital motions, $\sigma=\sqrt{gk}$ is the angular frequency, and $D_x = \frac{1}{2}\rho C_D \langle a \rangle_y P_x |u_x|u_x$ is the wave drag force on the canopy with $\langle a \rangle_y$ being the lateral-averaged FAD. Substituting equation \eqref{eq:wave_dissipation} into equation \eqref{eq:wave_energy} yields,
\begin{equation} \label{eq:wave_amp}
    \frac{1}{2} g c_g \frac{\partial a_w^2}{\partial x} = -G a_w^3, 
\end{equation}
in which 
\begin{equation} \label{eq:coeff_G}
    G = \frac{4}{3}\alpha_D C_D P_x \sigma^3 \int_{-h_b}^{0} a e^{3kz} \ \mathrm{d} z, 
\end{equation}
The solution of \eqref{eq:wave_amp} is
\begin{equation}
    \frac{a_{w}}{a_w^0} = \left( {1+\frac{Ga_{w}^0}{gc_g}x} \right)^{-1}, 
\end{equation}
in which $a_{w}^0$ (=0.8 m here) is the incident wave amplitude before entering the macroalage canopy. From the values of parameters reported above, the wave height decay over a 800-m (400-m) long farm is about 2.9\% (1.4\%), and the corresponding decay in Stokes drift velocity is 5.7\% (2.8\%).

\bibliographystyle{jfm}
\bibliography{jfm-instructions}

\begin{thebibliography}{77}
\expandafter\ifx\csname natexlab\endcsname\relax\def\natexlab#1{#1}\fi
\def\au#1{#1} \def\ed#1{#1} \def\yr#1{#1}\def\at#1{#1}\def\jt#1{\textit{#1}}
  \def\bt#1{#1}\def\bvol#1{\textbf{#1}} \def\vol#1{#1} \def\pg#1{#1}
  \def\publ#1{#1}\def\arxiv#1{#1}\def\org#1{#1}\def\st#1{\textit{#1}}

\bibitem[Abdolahpour {\em et~al.\/}(2017)Abdolahpour, Hambleton \&
  Ghisalberti]{Abdolahpour:2017JGR}
{\sc \au{Abdolahpour, M.}, \au{Hambleton, M.} \& \au{Ghisalberti, M.}}
  \yr{2017}  \at{The wave‐driven current in coastal canopies.}  \jt{J.
  Geophys. Res. Oceans}  \bvol{122},  \pg{3660--3674}.

\bibitem[Akselsen \& Ellingsen(2019)]{akselsen2019JFM}
{\sc \au{Akselsen, A.~H.} \& \au{Ellingsen, S.~Å.}} \yr{2019}  \at{Weakly
  nonlinear transient waves on a shear current: ring waves and skewed langmuir
  rolls}.  \jt{Journal of Fluid Mechanics}  \bvol{863},  \pg{114–149}.

\bibitem[Akselsen \& Ellingsen(2020)]{akselsen2020JFM}
{\sc \au{Akselsen, A.~H.} \& \au{Ellingsen, S.~Å.}} \yr{2020}
  \at{Langmuir-type vortices in wall-bounded flows driven by a criss-cross wavy
  wall topography}.  \jt{J. Fluid Mech.}  \bvol{900},  \pg{A19}.

\bibitem[Aylor \& Flesch(2001)]{Aylor:2001JAM}
{\sc \au{Aylor, D.~E.} \& \au{Flesch, T.~K.}} \yr{2001}  \at{Estimating spore
  release rates using a lagrangian stochastic simulation model.}  \jt{J. Appl.
  Meteorol.}  \bvol{40},  \pg{1196--1208}.

\bibitem[Barton {\em et~al.\/}(2014)Barton, Ward, Williams \&
  Follows]{Barton2014}
{\sc \au{Barton, A.~D.}, \au{Ward, B.~A.}, \au{Williams, R.~G.} \& \au{Follows,
  M.~J.}} \yr{2014}  \at{The impact of fine-scale turbulence on phytoplankton
  community structure}.  \jt{Limnol. Oceanogr. Fluids Environ.}  \bvol{4}~(1),
  \pg{34--49}.

\bibitem[Belcher {\em et~al.\/}(2003)Belcher, Jerram \& Hunt]{Belcher2003jfm}
{\sc \au{Belcher, S.~E.}, \au{Jerram, N.} \& \au{Hunt, J.~C.~R.}} \yr{2003}
  \at{Adjustment of a turbulent boundary layer to a canopy of roughness
  elements}.  \jt{J. Fluid Mech.}  \bvol{488},  \pg{369–398}.

\bibitem[Belcher(2012)]{Belcher2012grl}
{\sc \au{Belcher, S. E. et~al.}} \yr{2012}  \at{A global perspective on
  langmuir turbulence in the ocean surface boundary layer}.  \jt{Geophys. Res.
  Lett.}  \bvol{39}~(18),  \pg{L18605}.

\bibitem[Boller \& Carrington(2006)]{Boller2006jeb}
{\sc \au{Boller, M.~L.} \& \au{Carrington, E.}} \yr{2006}  \at{The hydrodynamic
  effects of shape and size change during reconfiguration of a flexible
  macroalga}.  \jt{J. Exp. Biol.}  \bvol{209}~(10),  \pg{1894--1903}.

\bibitem[Bou-Zeid {\em et~al.\/}(2005)Bou-Zeid, Meneveau \&
  Parlange]{Bou-Zeid:2005pof}
{\sc \au{Bou-Zeid, E.}, \au{Meneveau, C.} \& \au{Parlange, M.~B.}} \yr{2005}
  \at{A scale-dependent lagrangian dynamic model for large eddy simulation of
  complex turbulent flows.}  \jt{Phys. Fluids}  \bvol{17},  \pg{025105}.

\bibitem[Bradshaw(1987)]{Bradshaw1987ARFM}
{\sc \au{Bradshaw, P}} \yr{1987}  \at{Turbulent secondary flows}.  \jt{Annu.
  Rev. Fluid Mech.}  \bvol{19}~(1),  \pg{53--74}.

\bibitem[Cescatti \& Marcolla(2004)]{Cescatti2004AFM}
{\sc \au{Cescatti, A.} \& \au{Marcolla, B.}} \yr{2004}  \at{Drag coefficient
  and turbulence intensity in conifer canopies}.  \jt{Agric. For. Meteorol.}
  \bvol{121}~(3),  \pg{197--206}.

\bibitem[Chamecki {\em et~al.\/}(2019)Chamecki, Chor, Yang \&
  Meneveau]{Chamecki2019RoG}
{\sc \au{Chamecki, M.}, \au{Chor, T.}, \au{Yang, D.} \& \au{Meneveau, C.}}
  \yr{2019}  \at{Material transport in the ocean mixed layer: Recent
  developments enabled by large eddy simulations}.  \jt{Rev. Geophys.}
  \bvol{57}~(4),  \pg{1338--1371}.

\bibitem[Charrier {\em et~al.\/}(2018)Charrier, Wichard \&
  Reddy]{Charrier:2018}
{\sc \au{Charrier, B.}, \au{Wichard, T.} \& \au{Reddy, C.~R.~K.}} \yr{2018}
  {\em Protocols for Macroalgae Research\/}.  \publ{CRC Press}.

\bibitem[Chen {\em et~al.\/}(2019)Chen, Liu \& Zou]{Chen2019AWR}
{\sc \au{Chen, H.}, \au{Liu, X.} \& \au{Zou, Q.}} \yr{2019}  \at{Wave-driven
  flow induced by suspended and submerged canopies}.  \jt{Adv. Water Resour.}
  \bvol{123},  \pg{160--172}.

\bibitem[Craik(1977)]{Craik:1977JFM}
{\sc \au{Craik, A. D.~D.}} \yr{1977}  \at{The generation of langmuir
  circulations by an instability mechanism}.  \jt{J. Fluid Mech.}
  \bvol{81}~(2),  \pg{209–223}.

\bibitem[Craik \& Leibovich(1976)]{Craik:1976JFM}
{\sc \au{Craik, A. D.~D.} \& \au{Leibovich, S.}} \yr{1976}  \at{A rational
  model for langmuir circulations}.  \jt{J. Fluid Mech.}  \bvol{73}~(3),
  \pg{401–426}.

\bibitem[Dalrymple {\em et~al.\/}(1984)Dalrymple, Kirby \&
  Hwang]{Dalrymple1984}
{\sc \au{Dalrymple, R.~A.}, \au{Kirby, J.~T.} \& \au{Hwang, P.~A.}} \yr{1984}
  \at{Wave diffraction due to areas of energy dissipation}.  \jt{J. Water.
  Port, Coastal, Ocean Eng.}  \bvol{110}~(1),  \pg{67--79}.

\bibitem[D'Asaro(2001)]{D'Asaro2001JPO}
{\sc \au{D'Asaro, E.~A.}} \yr{2001}  \at{Turbulent vertical kinetic energy in
  the ocean mixed layer}.  \jt{J. Phys. Oceanogr.}  \bvol{31}~(12),
  \pg{3530--3537}.

\bibitem[Dayton(1985)]{Dayton:1985ARES}
{\sc \au{Dayton, P.~K.}} \yr{1985}  \at{Ecology of kelp communities.}
  \jt{Annu. Rev. Ecol. Syst.}  \bvol{16},  \pg{215--245}.

\bibitem[Duggins {\em et~al.\/}(1990)Duggins, Eckman \&
  Sewell]{Duggins:1990jembe}
{\sc \au{Duggins, D.~O.}, \au{Eckman, J.~E.} \& \au{Sewell, A.~T.}} \yr{1990}
  \at{Ecology of understory kelp environments. \uppercase{II}.
  \uppercase{E}ffects of kelps on recruitment of benthic invertebrates.}
  \jt{J. Exp. Mar. Biol. Ecol.}  \bvol{143},  \pg{27--45}.

\bibitem[Dupont \& Brunet(2008)]{Dupont:2008AFM}
{\sc \au{Dupont, S.} \& \au{Brunet, Y.}} \yr{2008}  \at{Influence of foliar
  density profile on canopy flow: A large-eddy simulation study.}  \jt{Agric.
  For. Meteorol.}  \bvol{148},  \pg{976--990}.

\bibitem[Finnigan {\em et~al.\/}(2009)Finnigan, Shaw \&
  Patton]{Finnigan2009JFM}
{\sc \au{Finnigan, J.~J.}, \au{Shaw, R.~H.} \& \au{Patton, E}} \yr{2009}
  \at{Turbulence structure above a vegetation canopy.}  \jt{J. Fluid Mech.}
  \bvol{637},  \pg{387--424}.

\bibitem[Fram {\em et~al.\/}(2008)Fram, Stewart, Brzezinski, Gaylord, Reed,
  Williams \& MacIntyre]{Fram2008LO}
{\sc \au{Fram, J.~P.}, \au{Stewart, H.~L.}, \au{Brzezinski, M.~A.},
  \au{Gaylord, B.}, \au{Reed, D.~C.}, \au{Williams, S.~L.} \& \au{MacIntyre,
  S.}} \yr{2008}  \at{Physical pathways and utilization of nitrate supply to
  the giant kelp, macrocystis pyrifera}.  \jt{Limnol. Oceanogr.}
  \bvol{53}~(4),  \pg{1589--1603}.

\bibitem[Gaylord {\em et~al.\/}(2007)Gaylord, Rosman, Reed, Koseff, MacIntyre,
  Arkema, McDonald, Brzezinski, Largier, Monismith, Raimondi \&
  Mardian]{Gaylord:2007lo}
{\sc \au{Gaylord, B.}, \au{Rosman, J.~H.}, \au{Reed, D.~C.}, \au{Koseff,
  J.~R.and~Fram, J.}, \au{MacIntyre, S.}, \au{Arkema, K.}, \au{McDonald, C.},
  \au{Brzezinski, M.~A.}, \au{Largier, J.~L.}, \au{Monismith, S.~G.},
  \au{Raimondi, P.~T.} \& \au{Mardian, B.}} \yr{2007}  \at{Spatial patterns of
  flow and their modification within and around a giant kelp forest.}
  \jt{Limnol. Oceanogr.}  \bvol{52},  \pg{1838--1852}.

\bibitem[Grant \& Belcher(2009)]{Grant2009JPO}
{\sc \au{Grant, A. L.~M.} \& \au{Belcher, S.~E.}} \yr{2009}
  \at{Characteristics of langmuir turbulence in the ocean mixed layer}.  \jt{J.
  Phys. Oceanogr.}  \bvol{39}~(8),  \pg{1871--1887}.

\bibitem[Harcourt \& D’Asaro(2008)]{Harcourt2008JPO}
{\sc \au{Harcourt, R.~R.} \& \au{D’Asaro, E.~A.}} \yr{2008}  \at{Large-eddy
  simulation of langmuir turbulence in pure wind seas}.  \jt{J. Phys.
  Oceanogr.}  \bvol{38}~(7),  \pg{1542--1562}.

\bibitem[Henderson(2019)]{Henderson2019ce}
{\sc \au{Henderson, S.~M.}} \yr{2019}  \at{Motion of buoyant, flexible aquatic
  vegetation under waves: Simple theoretical models and parameterization of
  wave dissipation}.  \jt{Coast. Eng.}  \bvol{152},  \pg{103497}.

\bibitem[Huai {\em et~al.\/}(2012)Huai, Hu, Zeng \& Han]{Huai2012awr}
{\sc \au{Huai, W.~X.}, \au{Hu, Y.}, \au{Zeng, Y.~H.} \& \au{Han, J.}} \yr{2012}
   \at{Velocity distribution for open channel flows with suspended vegetation}.
   \jt{Adv. Water Resour.}  \bvol{49},  \pg{56–61}.

\bibitem[Koehl \& Wainwright(1977)]{Koehl:1977lo}
{\sc \au{Koehl, M. A.~R.} \& \au{Wainwright, S.~A.}} \yr{1977}  \at{Mechanical
  adaptations of a giant kelp.}  \jt{Limnol. Oceanogr.}  \bvol{22},
  \pg{1067--1071}.

\bibitem[Kukulka {\em et~al.\/}(2010)Kukulka, Plueddemann, Trowbridge \&
  Sullivan]{Kukulka2010JPO}
{\sc \au{Kukulka, T.}, \au{Plueddemann, A.~J.}, \au{Trowbridge, J.~H.} \&
  \au{Sullivan, P.~P.}} \yr{2010}  \at{Rapid mixed layer deepening by the
  combination of langmuir and shear instabilities: A case study}.  \jt{J. Phys.
  Oceanogr.}  \bvol{40}~(11),  \pg{2381--2400}.

\bibitem[Legg \& Powell(1979)]{LEGG1979}
{\sc \au{Legg, B.~J.} \& \au{Powell, F.~A.}} \yr{1979}  \at{Spore dispersal in
  a barley crop: A mathematical model}.  \jt{Agric. Meteorol.}  \bvol{20}~(1),
  \pg{47--67}.

\bibitem[Leibovich(1977)]{Leibovich1977jfm}
{\sc \au{Leibovich, S.}} \yr{1977}  \at{Convective instability of stably
  stratified water in the ocean}.  \jt{J. Fluid Mech.}  \bvol{82}~(3),
  \pg{561--581}.

\bibitem[Leibovich(1983)]{Leibovich1983arfm}
{\sc \au{Leibovich, S.}} \yr{1983}  \at{The form and dynamics of langmuir
  circulations}.  \jt{Annu. Rev. Fluid Mech.}  \bvol{15}~(1),  \pg{391--427}.

\bibitem[Luhar {\em et~al.\/}(2010)Luhar, Coutu, Infantes, Fox \&
  Nepf]{Luhar2010JGR}
{\sc \au{Luhar, M.}, \au{Coutu, S.}, \au{Infantes, E.}, \au{Fox, S.} \&
  \au{Nepf, H.}} \yr{2010}  \at{Wave-induced velocities inside a model seagrass
  bed}.  \jt{J. Geophys. Res. Oceans}  \bvol{115}~(C12),  \pg{C12005}.

\bibitem[Luhar {\em et~al.\/}(2013)Luhar, Infantes, Orfila, Terrados \&
  Nepf]{Luhar2013JGR}
{\sc \au{Luhar, M.}, \au{Infantes, E.}, \au{Orfila, A.}, \au{Terrados, J.} \&
  \au{Nepf, H.}} \yr{2013}  \at{Field observations of wave‐induced streaming
  through a submerged seagrass (posidonia oceanica) meadow}.  \jt{J. Geophys.
  Res. Oceans}  \bvol{118},  \pg{1955–1968}.

\bibitem[Luhar \& Nepf(2011)]{Luhar2011lo}
{\sc \au{Luhar, M.} \& \au{Nepf, H.~M.}} \yr{2011}  \at{Flow‐induced
  reconfiguration of buoyant and flexible aquatic vegetation}.  \jt{Limnol.
  Oceanogr.}  \bvol{56}~(6),  \pg{2003--2017}.

\bibitem[Marcolla {\em et~al.\/}(2003)Marcolla, Pitacco \&
  Cescatti]{Marcolla2003BLM}
{\sc \au{Marcolla, B.}, \au{Pitacco, A.} \& \au{Cescatti, A.}} \yr{2003}
  \at{Canopy architecture and turbulence structure in a coniferous forest}.
  \jt{Boundary-Layer Meteorol.}  \bvol{108},  \pg{39--59}.

\bibitem[McWiliams {\em et~al.\/}(1997)McWiliams, Sullivan \&
  Moeng]{McWiliams:1997jfm}
{\sc \au{McWiliams, J.}, \au{Sullivan, P.} \& \au{Moeng, C.}} \yr{1997}
  \at{Langmuir turbulence in the ocean.}  \jt{J. Fluid Mech.}  \bvol{334},
  \pg{1--30}.

\bibitem[McWilliams(2006)]{McWilliams2006}
{\sc \au{McWilliams, J.}} \yr{2006} {\em {F}undamentals of {G}eophysical
  {F}luid {D}ynamics\/}.  \publ{Cambridge University Press}.

\bibitem[Monismith \& Fong(2004)]{Monismith:2004lo}
{\sc \au{Monismith, S.~G.} \& \au{Fong, D.~A.}} \yr{2004}  \at{A note on the
  potential transport of scalars and organisms by surface waves.}  \jt{Limnol.
  Oceanogr.}  \bvol{49},  \pg{1214--1217}.

\bibitem[Nepf(2012{\natexlab{{\em a\/}}})]{Nepf:2012ARFM}
{\sc \au{Nepf, H.~M.}} \yr{2012{\natexlab{{\em a\/}}}}  \at{Flow and transport
  in regions with aquatic vegetation.}  \jt{Annu. Rev. Fluid Mech.}  \bvol{44},
   \pg{123--142}.

\bibitem[Nepf(2012{\natexlab{{\em b\/}}})]{Nepf:2012JHR}
{\sc \au{Nepf, H.~M.}} \yr{2012{\natexlab{{\em b\/}}}}  \at{Hydrodynamics of
  vegetated channels.}  \jt{J. Hydraul. Res.}  \bvol{50},  \pg{262--279}.

\bibitem[Pan {\em et~al.\/}(2014)Pan, Chamecki \& Isard]{Pan2014jfm}
{\sc \au{Pan, Y.}, \au{Chamecki, M.} \& \au{Isard, S.~A.}} \yr{2014}
  \at{Large-eddy simulation of turbulence and particle dispersion inside the
  canopy roughness sublayer}.  \jt{J. Fluid Mech.}  \bvol{753},  \pg{499--534}.

\bibitem[Pan {\em et~al.\/}(2016)Pan, Chamecki \& Nepf]{Pan2016BLM}
{\sc \au{Pan, Y.}, \au{Chamecki, M.} \& \au{Nepf, H.~M.}} \yr{2016}
  \at{Estimating the instantaneous drag–wind relationship for a horizontally
  homogeneous canopy}.  \jt{Boundary-Layer Meteorol.}  \bvol{160},
  \pg{63--82}.

\bibitem[Pinard \& Wilson(2001)]{Pinard2001JAM}
{\sc \au{Pinard, J.~D.} \& \au{Wilson, J.~D.}} \yr{2001}  \at{First- and
  second-order closure models for wind in a plant canopy}.  \jt{J. Appl.
  Meteorol.}  \bvol{40}~(10),  \pg{1762--1768}.

\bibitem[Plew(2011{\natexlab{{\em a\/}}})]{Plew:2011jhe}
{\sc \au{Plew, D.~R.}} \yr{2011{\natexlab{{\em a\/}}}}  \at{Depth-averaged drag
  coefficient for modeling flow through suspended canopies.}  \jt{J. Hydraul.
  Eng.}  \bvol{137},  \pg{234–247}.

\bibitem[Plew(2011{\natexlab{{\em b\/}}})]{Plew:2011aei}
{\sc \au{Plew, D.~R.}} \yr{2011{\natexlab{{\em b\/}}}}  \at{Shellfish
  farm-induced changes to tidal circulation in an embayment, and implications
  for seston depletion.}  \jt{Aquacult. Env. Interac.}  \bvol{1},
  \pg{201–214}.

\bibitem[Plew {\em et~al.\/}(2006)Plew, Spigel, Stevens, Nokes \&
  Davidson]{Plew:2006efm}
{\sc \au{Plew, D.~R.}, \au{Spigel, R.~H.}, \au{Stevens, C.~L.}, \au{Nokes,
  R.~I.} \& \au{Davidson, M.~J.}} \yr{2006}  \at{Stratified flow interactions
  with a suspended canopy.}  \jt{Environ. Fluid Mech.}  \bvol{6},
  \pg{519--539}.

\bibitem[Plew {\em et~al.\/}(2005)Plew, Stevens, Spigel \& D.]{Plew:2005joe}
{\sc \au{Plew, D.~R.}, \au{Stevens, C.~L.}, \au{Spigel, R.~H.} \& \au{D.,
  Hartstein~N.}} \yr{2005}  \at{Hydrodynamic implications of large offshore
  mussel farms.}  \jt{IEEE J. Ocean Eng.}  \bvol{30},  \pg{95–108}.

\bibitem[Polton {\em et~al.\/}(2008)Polton, Smith, MacKinnon \&
  Tejada-Mart{\'i}nez]{Polton2008GRL}
{\sc \au{Polton, J.~A.}, \au{Smith, J.~A.}, \au{MacKinnon, J.~A.} \&
  \au{Tejada-Mart{\'i}nez, A.~E.}} \yr{2008}  \at{Rapid generation of
  high-frequency internal waves beneath a wind and wave forced oceanic surface
  mixed layer}.  \jt{Geophys. Res. Lett.}  \bvol{35}~(13),  \pg{L13602}.

\bibitem[Pope(2000)]{Pope2000}
{\sc \au{Pope, Stephen~B.}} \yr{2000} {\em Turbulent Flows\/}.  \publ{Cambridge
  University Press}.

\bibitem[Rominger \& Nepf(2011)]{Rominger2011JFM}
{\sc \au{Rominger, J.~T.} \& \au{Nepf, H.~M.}} \yr{2011}  \at{Flow adjustment
  and interior flow associated with a rectangular porous obstruction}.  \jt{J.
  Fluid Mech.}  \bvol{680},  \pg{636–659}.

\bibitem[van Rooijen {\em et~al.\/}(2020)van Rooijen, Lowe, Rijnsdorp,
  Ghisalberti, Jacobsen \& McCall]{Rooijen2020JGR}
{\sc \au{van Rooijen, A.}, \au{Lowe, R.}, \au{Rijnsdorp, D.}, \au{Ghisalberti,
  M.}, \au{Jacobsen, N.~G.} \& \au{McCall, R.}} \yr{2020}  \at{Wave‐driven
  mean flow dynamics in submerged canopies.}  \jt{J. Geophys. Res. Oceans}
  \bvol{125},  \pg{e2019JC015935}.

\bibitem[Rosman {\em et~al.\/}(2013)Rosman, Denny, Zeller, Monismith \&
  Koseff]{Rosman:2013lo}
{\sc \au{Rosman, J.~H.}, \au{Denny, M.~W.}, \au{Zeller, R.~B.}, \au{Monismith,
  S.~G.} \& \au{Koseff, J.~R.}} \yr{2013}  \at{Interaction of waves and
  currents with kelp forests (macrocystis pyrifera): Insights from a
  dynamically scaled laboratory model.}  \jt{Limnol. Oceanogr.}  \bvol{58},
  \pg{790--802}.

\bibitem[Rosman {\em et~al.\/}(2007)Rosman, Koseff, Monismith \&
  Grover]{Rosman:2007JGR}
{\sc \au{Rosman, J.~H.}, \au{Koseff, J.~R.}, \au{Monismith, S.~G.} \&
  \au{Grover, J.}} \yr{2007}  \at{A field investigation into the effects of a
  kelp forest (macrocystis pyrifera) on coastal hydrodynamics and transport.}
  \jt{J. Geophys. Res.}  \bvol{112},  \pg{C02016}.

\bibitem[Rosman {\em et~al.\/}(2010)Rosman, Monismith, Denny \&
  Koseff]{Rosman2010lo}
{\sc \au{Rosman, Johanna~H.}, \au{Monismith, Stephen~G.}, \au{Denny, Mark~W.}
  \& \au{Koseff, Jeffrey~R.}} \yr{2010}  \at{Currents and turbulence within a
  kelp forest (macrocystis pyrifera): Insights from a dynamically scaled
  laboratory model}.  \jt{Limnol. Oceanogr.}  \bvol{55}~(3),  \pg{1145--1158}.

\bibitem[Schiel \& Forster(2015)]{Schiel:2015}
{\sc \au{Schiel, D.~R.} \& \au{Forster, M.~S.}} \yr{2015} {\em The biology and
  ecology of giant kelp forests\/}.  \publ{University of California Press}.

\bibitem[Shaw \& Schumann(1992)]{Shaw:1992BLM}
{\sc \au{Shaw, R.~H.} \& \au{Schumann, U.}} \yr{1992}  \at{Large-eddy
  simulation of turbulent flow above and within a forest.}  \jt{Boundary-Layer
  Meteorol.}  \bvol{61},  \pg{47–64}.

\bibitem[Shrestha \& Anderson(2019)]{Shrestha2019EFM}
{\sc \au{Shrestha, K.} \& \au{Anderson, W.}} \yr{2019}  \at{Coastal langmuir
  circulations induce phase-locked modulation of bathymetric stress}.
  \jt{Environ. Fluid Mech.}  \bvol{20},  \pg{873–884}.

\bibitem[Skitka {\em et~al.\/}(2020)Skitka, Marston \&
  Fox-Kemper]{Skitka2020JPO}
{\sc \au{Skitka, J.}, \au{Marston, J.~B.} \& \au{Fox-Kemper, B.}} \yr{2020}
  \at{{Reduced-Order Quasilinear Model of Ocean Boundary-Layer Turbulence}}.
  \jt{J. Phys. Oceanogr.}  \bvol{50}~(3),  \pg{537--558}.

\bibitem[Skyllingstad \& Denbo(1995)]{Skyllingstad:1995jgr}
{\sc \au{Skyllingstad, E.~D.} \& \au{Denbo, D.~W.}} \yr{1995}  \at{An ocean
  large‐eddy simulation of langmuir circulations and convection in the
  surface mixed layer.}  \jt{J. Geophys. Res. Oceans}  \bvol{100},
  \pg{8501--8522}.

\bibitem[Stevens \& Plew(2019)]{Stevens2019fmars}
{\sc \au{Stevens, C.} \& \au{Plew, D.}} \yr{2019}  \at{Bridging the separation
  between studies of the biophysics of natural and built marine canopies}.
  \jt{Front. Mar. Sci.}  \bvol{6},  \pg{217}.

\bibitem[Stevens \& Petersen(2011)]{Stevens:2011aei}
{\sc \au{Stevens, C.~L.} \& \au{Petersen, J.~K.}} \yr{2011}  \at{Turbulent,
  stratified flow through a suspended shellfish canopy: implications for mussel
  farm design.}  \jt{Aquacult. Env. Interac.}  \bvol{36},  \pg{87--104}.

\bibitem[Stevens {\em et~al.\/}(2014)Stevens, Graham \&
  Meneveau]{Stevens2014RE}
{\sc \au{Stevens, R. J. A.~M.}, \au{Graham, J.} \& \au{Meneveau, C.}} \yr{2014}
   \at{A concurrent precursor inflow method for large eddy simulations and
  applications to finite length wind farms}.  \jt{Renew. Energy}  \bvol{68},
  \pg{46 -- 50}.

\bibitem[Sullivan {\em et~al.\/}(1998)Sullivan, Moeng, Stevens, Lenschow \&
  Mayor]{Sullivan:1998jas}
{\sc \au{Sullivan, P.~P.}, \au{Moeng, C.~H.}, \au{Stevens, B.}, \au{Lenschow,
  D.~H.} \& \au{Mayor, S.~D.}} \yr{1998}  \at{Structure of the entrainment zone
  capping the convective atmospheric boundary layer}.  \jt{J. Atmos. Sci.}
  \bvol{55},  \pg{3042--3064}.

\bibitem[Suzuki \& Fox-Kemper(2016)]{Suzuki2016JGR}
{\sc \au{Suzuki, N.} \& \au{Fox-Kemper, B.}} \yr{2016}  \at{Understanding
  stokes forces in the wave-averaged equations}.  \jt{J. Geophys. Res. Oceans}
  \bvol{121}~(5),  \pg{3579--3596}.

\bibitem[Taylor \& Sarkar(2008)]{Taylor2008b}
{\sc \au{Taylor, J.~R.} \& \au{Sarkar, S.}} \yr{2008}  \at{Stratification
  effects in a bottom ekman layer}.  \jt{J. Phys. Oceanogr.}  \bvol{38}~(11),
  \pg{2535--2555}.

\bibitem[Thorpe(2004)]{Thorpe2004ARFM}
{\sc \au{Thorpe, S.~A.}} \yr{2004}  \at{Langmuir circulation}.  \jt{Annu. Rev.
  Fluid Mech.}  \bvol{36}~(1),  \pg{55--79}.

\bibitem[Troell {\em et~al.\/}(2009)Troell, Joyce, Chopin, Neori, Buschmann \&
  Fang]{Troell:2009}
{\sc \au{Troell, M.}, \au{Joyce, A.}, \au{Chopin, T.}, \au{Neori, A.},
  \au{Buschmann, A.~H.} \& \au{Fang, J.~G.}} \yr{2009}  \at{Ecological
  engineering in aquaculture - potential for integrated multi-trophic
  aquaculture (\uppercase{IMTA}) in marine offshore systems.}  \jt{Aquaculture}
   \bvol{297},  \pg{1--9}.

\bibitem[Tseung {\em et~al.\/}(2016)Tseung, Kikkert \& Plew]{Tseung:2016efm}
{\sc \au{Tseung, H.~L.}, \au{Kikkert, G.~A.} \& \au{Plew, D.}} \yr{2016}
  \at{Hydrodynamics of suspended canopies with limited length and width.}
  \jt{Environ. Fluid Mech.}  \bvol{16},  \pg{145–166}.

\bibitem[Utter \& Denny(1996)]{Utter:1996}
{\sc \au{Utter, B.} \& \au{Denny, M.}} \yr{1996}  \at{Wave-induced forces on
  the giant kelp \uppercase{M}acrocystis pyrifera (\uppercase{A}gardh): field
  test of a computational model.}  \jt{J. Exp. Biol.}  \bvol{199},
  \pg{2645–2654}.

\bibitem[Van~Roekel {\em et~al.\/}(2012)Van~Roekel, Fox-Kemper, Sullivan,
  Hamlington \& Haney]{Roekel2012JGR}
{\sc \au{Van~Roekel, L.~P.}, \au{Fox-Kemper, B.}, \au{Sullivan, P.~P.},
  \au{Hamlington, P.~E.} \& \au{Haney, S.~R.}} \yr{2012}  \at{The form and
  orientation of langmuir cells for misaligned winds and waves}.  \jt{J.
  Geophys. Res. Oceans}  \bvol{117}~(C5).

\bibitem[Vogel(1989)]{Vogel:1989}
{\sc \au{Vogel, S.}} \yr{1989}  \at{Drag and reconfiguration of broad leaves in
  high winds.}  \jt{J. Exp. Bot.}  \bvol{40},  \pg{941–948}.

\bibitem[Yan {\em et~al.\/}(2017)Yan, Nepf, Huang \& Cui]{Yan:2017efm}
{\sc \au{Yan, C.}, \au{Nepf, H.~M.}, \au{Huang, W.} \& \au{Cui, G.}} \yr{2017}
  \at{Large eddy simulation of flow and scalar transport in a vegetated
  channel.}  \jt{Environ. Fluid Mech.}  \bvol{17},  \pg{497–519}.

\bibitem[Yang {\em et~al.\/}(2015)Yang, Chen, Chamecki \&
  Meneveau]{Yang:2015jgr}
{\sc \au{Yang, D.}, \au{Chen, B.}, \au{Chamecki, M.} \& \au{Meneveau, C.}}
  \yr{2015}  \at{Oil plumes and dispersion in langmuir, upper‐ocean
  turbulence: Large‐eddy simulations and k‐profile parameterization.}
  \jt{J. Geophys. Res. Oceans}  \bvol{120},  \pg{4729–4759}.

\bibitem[Zhao {\em et~al.\/}(2017)Zhao, X. \& Li]{Zhao2017AWR}
{\sc \au{Zhao, F.}, \au{X., Huai~W.} \& \au{Li, D.}} \yr{2017}  \at{Numerical
  modeling of open channel flow with suspended canopy}.  \jt{Adv. Water
  Resour.}  \bvol{105},  \pg{132–143}.

\bibitem[Zhou \& Venayagamoorthy(2019)]{Zhou:2019jfm}
{\sc \au{Zhou, J.} \& \au{Venayagamoorthy, S.}} \yr{2019}  \at{Near-field mean
  flow dynamics of a cylindrical canopy patch suspended in deep water}.  \jt{J.
  Fluid Mech.}  \bvol{858},  \pg{634–655}.

\end{thebibliography}

\end{document}